\documentclass[preprint,twocolumn]{aastex6}
\newcommand{\taraa}{K2-117}
\newcommand{\tarab}{K2-146}
\newcommand{\tarad}{K2-122}
\newcommand{\tarac}{K2-123}
\newcommand{\tarba}{K2-147}
\newcommand{\tarca}{EPIC 220187552}
\newcommand{\tarcb}{EPIC 220194953}
\newcommand{\tarcc}{K2-148}
\newcommand{\tarcd}{K2-149}
\newcommand{\tarce}{K2-150}
\newcommand{\tarcf}{K2-151}
\newcommand{\tarda}{K2-152}
\newcommand{\tardb}{K2-153}
\newcommand{\tardc}{K2-154}
\usepackage{amsmath,amssymb}
\usepackage{natbib}
\usepackage{aas_macros} 
\slugcomment{}
\shortauthors{Hirano et al.}
\shorttitle{Validation of M-dwarf Planets in {\it K2} Campaign Fields 5 -- 10}
\begin{document}
\title{Exoplanets around Low-mass Stars Unveiled by K2}
\author{
Teruyuki Hirano\altaffilmark{1}, 
Fei Dai\altaffilmark{2,3},
Davide Gandolfi\altaffilmark{4},
Akihiko Fukui\altaffilmark{5},
John H. Livingston\altaffilmark{6},
Kohei Miyakawa\altaffilmark{1},
Michael Endl\altaffilmark{7},
William D. Cochran\altaffilmark{7},
Francisco J. Alonso-Floriano\altaffilmark{8,9},
Masayuki Kuzuhara\altaffilmark{10,11},
David Montes\altaffilmark{9},
Tsuguru Ryu\altaffilmark{12,11},
Simon Albrecht\altaffilmark{13},
Oscar Barragan\altaffilmark{4},
Juan Cabrera\altaffilmark{14},
Szilard Csizmadia\altaffilmark{14},
Hans Deeg\altaffilmark{15,16},
Philipp Eigm\"uller\altaffilmark{14},
Anders Erikson\altaffilmark{14},
Malcolm Fridlund\altaffilmark{8,17},
Sascha Grziwa\altaffilmark{18},
Eike W. Guenther\altaffilmark{19},
Artie P. Hatzes\altaffilmark{19},
Judith Korth\altaffilmark{18},
Tomoyuki Kudo\altaffilmark{20},
Nobuhiko Kusakabe\altaffilmark{10,11},
Norio Narita\altaffilmark{6,10,11},
David Nespral\altaffilmark{15,16},
Grzegorz Nowak\altaffilmark{15,16},
Martin P\"atzold\altaffilmark{18},
Enric Palle\altaffilmark{15,16},
Carina M. Persson\altaffilmark{17},
Jorge Prieto-Arranz\altaffilmark{15,16},
Heike Rauer\altaffilmark{14,21},
Ignasi Ribas\altaffilmark{22},
Bun'ei Sato\altaffilmark{1},
Alexis M. S. Smith\altaffilmark{14},
Motohide Tamura\altaffilmark{6,10,11},
Yusuke Tanaka\altaffilmark{6},
Vincent Van Eylen\altaffilmark{8},
Joshua N.\ Winn\altaffilmark{3}
} 
\altaffiltext{1}{Department of Earth and Planetary Sciences, Tokyo Institute of Technology,
2-12-1 Ookayama, Meguro-ku, Tokyo 152-8551, Japan}
\email{hirano@geo.titech.ac.jp}
\altaffiltext{2}{Department of Physics, and Kavli Institute for Astrophysics and Space Research, Massachusetts Institute of Technology, Cambridge, MA 02139, USA}
\altaffiltext{3}{Department of Astrophysical Sciences, Princeton University, 4 Ivy Lane, Princeton, NJ 08544, USA}
\altaffiltext{4}{Dipartimento di Fisica, Universit\'a di Torino, via P. Giuria 1, 10125 Torino, Italy}
\altaffiltext{5}{Okayama Astrophysical Observatory, National Astronomical Observatory of Japan, Asakuchi, Okayama 719-0232, Japan}
\altaffiltext{6}{Department of Astronomy, Graduate School of Science, The University of Tokyo, Hongo 7-3-1, Bunkyo-ku, Tokyo, 113-0033, Japan}
\altaffiltext{7}{Department of Astronomy and McDonald Observatory, University of Texas at Austin, 2515 Speedway,~Stop~C1400,~Austin,~TX~78712,~USA}
\altaffiltext{8}{Leiden Observatory, Leiden University, 2333CA Leiden, The Netherlands}
\altaffiltext{9}{Departamento de Astrof\'isica y Ciencias de la Atm\'osfera, Facultad de Ciencias F\'isicas, Universidad Complutense de Madrid, 28040 Madrid, Spain}
\altaffiltext{10}{Astrobiology Center, NINS, 2-21-1 Osawa, Mitaka, Tokyo 181-8588, Japan}
\altaffiltext{11}{National Astronomical Observatory of Japan, NINS, 2-21-1 Osawa, Mitaka, Tokyo 181-8588, Japan}
\altaffiltext{12}{SOKENDAI (The Graduate University for Advanced Studies), 2-21-1 Osawa, Mitaka, Tokyo 181-8588, Japan}
\altaffiltext{13}{Stellar Astrophysics Centre, Department of Physics and Astronomy, Aarhus University, Ny Munkegade 120, DK-8000 Aarhus C, Denmark}
\altaffiltext{14}{Institute of Planetary Research, German Aerospace Center, Rutherfordstrasse 2, 12489 Berlin, Germany}
\altaffiltext{15}{Instituto de Astrof\'\i sica de Canarias, C/\,V\'\i a L\'actea s/n, 38205 La Laguna, Spain}
\altaffiltext{16}{Departamento de Astrof\'isica, Universidad de La Laguna, 38206 La Laguna, Spain}
\altaffiltext{17}{Department of Space, Earth and Environment, Chalmers University of Technology, Onsala Space Observatory, 439 92 Onsala, Sweden}
\altaffiltext{18}{Rheinisches Institut f\"ur Umweltforschung an der Universit\"at zu K\"oln, Aachener Strasse 209, 50931 K\"oln, Germany}
\altaffiltext{19}{Th\"uringer Landessternwarte Tautenburg, Sternwarte 5, D-07778 Tautenberg, Germany}
\altaffiltext{20}{Subaru Telescope, National Astronomical Observatory of Japan, 650 North Aohoku Place, Hilo, HI 96720, USA}
\altaffiltext{21}{Center for Astronomy and Astrophysics, TU Berlin, Hardenbergstr. 36, 10623 Berlin, Germany}
\altaffiltext{22}{Institut de Ci\`{e}ncies de l'Espai (CSIC-IEEC), Carrer de Can Magrans, Campus UAB, 08193 Bellaterra, Spain}

\begin{abstract}
We present the detection and follow-up observations of planetary
candidates around low-mass stars observed by the {\it K2}
mission. Based on light-curve analysis, adaptive-optics imaging, and
optical spectroscopy at low and high resolution (including radial
velocity measurements), we validate 16 planets around 12 low-mass
stars observed during {\it K2} campaigns 5--10. Among the 16 planets,
12 are newly validated, with orbital periods ranging from 0.96--33
days.  For one of the planets (\tarcf b) we present ground-based
transit photometry, allowing us to refine the ephemerides.  
Combining our {\it K2} M-dwarf planets together with the
validated or confirmed planets found previously, we investigate the
dependence of planet radius $R_p$ on stellar insolation and
metallicity [Fe/H].  We confirm that for periods $P\lesssim 2$ days,
planets with a radius $R_p\gtrsim 2\,R_\oplus$ are less common than
planets with a radius between 1--2\,$R_\oplus$.  We also see a hint of
the ``radius valley" between 1.5 and 2~$R_\oplus$ that has been seen
for close-in planets around FGK stars.  These features in the
radius/period distribution could be attributed to photoevaporation of
planetary envelopes by high-energy photons from the host star, as they
have for FGK stars.  For the M dwarfs, though, the features are not as
well defined, and we cannot rule out other explanations such as
atmospheric loss from internal planetary heat sources, or truncation
of the protoplanetary disk.  There also appears to be a relation
between planet size and metallicity: those few planets larger than
about 3~$R_\oplus$ are found around the most metal-rich M dwarfs.
\end{abstract}

\section{Introduction\label{s:intro}}\label{s:intro}

M dwarfs have some advantages over solar-type (FGK) stars in the detection and characterization of
transiting planets. Their smaller sizes lead to
deeper transits for a given planet radius. In addition, their habitable zones occur
at shorter orbital periods, facilitating the study of terrestrial planets in the habitable zone. 
These advantages are now widely appreciated.  Many observational and theoretical studies
have focused on M-dwarf planets, including their potential habitability and detectable biosignatures 
\citep[e.g.,][]{2007AsBio...7...85S, 2016PhR...663....1S}. 
However, the number of currently known transiting planets around low-mass stars is much smaller 
than that for solar-type stars, because low-mass stars are optically faint.
In particular, the number of mid-to-late M dwarfs ($T_\mathrm{eff}\lesssim 3500$\,K) hosting
transiting planets is extremely limited (fewer than 20, as of September 2017).
While the planets around early M dwarfs have been investigated in detail with the {\it Kepler}
sample \citep{2013ApJ...767...95D, 2015ApJ...807...45D, 2014ApJ...791...10M, 2015ApJ...798..112M, 
2015ApJ...814..130M, 2016ApJ...816...66B}, 
the distribution and properties of mid-to-late M-dwarf planetary systems are still relatively unexplored.

{\it Kepler}'s second mission, {\it K2} \citep{2014PASP..126..398H}, has also contributed to 
the search for transiting planets around M dwarfs. Hundreds of stars 
have been identified as candidate planet-hosting stars 
\citep[e.g.,][]{2015ApJ...809...25M, 2016ApJS..222...14V, 2016ApJS..226....7C, 2016MNRAS.461.3399P}, 
many of which have been validated
\citep[e.g.,][]{2017AJ....154..207D}. 
Moreover, {\it K2} has observed young stars in stellar clusters (e.g., the Hyades, Pleiades, and Beehive),
including many low-mass stars.
Several transiting planet candidates around these have already been reported
\citep{2016ApJ...818...46M, 2016AJ....152...61M, 2017AJ....153...64M, 2018AJ....155....4M, 2017arXiv170910398C}. 
These planets are potentially promising targets for follow-up studies such
as Doppler mass measurement and atmospheric characterization.

We have been participating in {\it K2} 
planet detection and characterization in the framework
of an international collaboration called KESPRINT\footnote{
In 2016, the two independent {\it K2} follow-up teams KEST (Kepler Exoplanet Science Team) and ESPRINT 
({\it Equipo de Seguimiento de Planetas Rocosos Intepretando sus Transitos}) merged 
and became the larger collaboration ``KESPRINT". 
}.
Making use of our own pipeline to reduce the {\it K2} data 
and look for transit signals, we have detected 30-80 planet candidates in each of the {\it K2} 
campaign fields. Through intensive follow-up observations using various facilities all over 
the world, we have validated or confirmed many transiting planets 
\citep[e.g.,][]{2015ApJ...812..112S, 2017A&A...604A..16F, 2017AJ....154..123G, 2017A&A...608A..93G}. 
In this paper, we focus on planetary systems around M dwarfs found by the KESPRINT project.

The rest of the paper is organized as follows. 
In Section \ref{s:k2}, we describe the reduction of the {\it K2} data and detection of 
the planet candidates by our pipeline. Next, we report our follow-up observations, 
including low- and high-resolution optical spectroscopy,
high-contrast imaging, and ground-based follow-up transit observations (Section \ref{s:obs}). 
Section \ref{s:analysis} presents the analysis of the follow-up observations,
through which we validate 15 planets around M dwarfs. 
Individual systems of special interest are described in Section \ref{s:individual}. 
In Section \ref{s:discussion} we examine the properties of all the transiting planets currently known around M dwarfs,
with a focus on the planetary radius.
Our conclusions are in Section \ref{s:conclusion}.

\section{{\it K2} Photometry and Detection of Planet Candidates\label{s:k2}}\label{s:k2}

\subsection{{\it K2} Light Curve Reduction\label{s:k2redution}}\label{s:k2reduction}
Due to the loss of two of its four reaction wheels, the {\it Kepler}
spacecraft can no longer maintain the pointing stability required
to observe its original field of view. The {\it Kepler} telescope was re-purposed for a new series
of observations under the name {\it K2} \citep{2014PASP..126..398H}. 
By observing in the ecliptic, the torque by solar radiation pressure is minimized, significantly
improving its pointing stability. The spacecraft must also
switch to a different field of view about every three months to maintain pointing away from the Sun.
In this operational mode, the photometry is strongly affected by the rolling motion of the spacecraft along its boresight
and the variation of pixel sensitivity. To reduce this effect, we
adopted an approach similar to that described by \citet{2014PASP..126..948V}. 

We now briefly describe our light-curve production pipeline. We
downloaded the target pixel files from the Mikulski Archive for Space
Telescopes.\footnote{https://archive.stsci.edu/k2.}  We then put down
circular apertures 
surrounding the brightest pixel within the collection of pixels recorded for each target.
We fitted a 2-D Gaussian function to the intensity distribution at each recorded time.
The resultant $X$ and $Y$ positions of the Gaussian
function, as a function of time, allowed us to track the rolling motion of the spacecraft. 
To reduce the intensity fluctuations associated with this motion,
we divided the apparent flux variation by the best-fitting piecewise linear
relationship between apparent flux and the coordinates $X$ and $Y$.
The systematic correction was described in more detail by \citet{2017AJ....153...40D}.

\subsection{Transit Detection}

To remove any long-term systematic 
or instrumental flux variations that may complicate the search of transit 
signals, we fitted the {\it K2} light curve with a cubic spline with 
a timescale of 1.5 days. The observed light curve was then divided by the spline fit.
The smoothing interval of 1.5 days was chosen to be much longer than the expected duration
of planetary transits, which are measured in hours for
for short-period planets around dwarf stars.
We then searched for periodic transit signals with the Box-Least-Squares
algorithm \citep{2002A&A...391..369K}. We employed a modification of the BLS algorithm, using a
more efficient nonlinear frequency grid that takes into account the
scaling of transit duration with orbital period \citep{2014A&A...561A.138O}. To
quantify the significance of a transit detection, we adopted the
signal detection efficiency (SDE) \citep{2014A&A...561A.138O} which is defined
by the amplitude of peak in the BLS spectrum normalized by the local
standard deviation.  A signal was considered significant if the
SDE is greater than 6.5. To search for any additional planets in the system, we re-computed
the BLS spectrum after removing the transit signal that was detected in the previous
iteration, until the maximum SDE dropped below 6.5.

\subsection{Initial Vetting}
After the transit signals were identified, we
performed a quick initial vetting process to exclude obvious false
positives. We sought evidence for any alternation in the eclipse depths or a
significant secondary eclipse, either of which would reveal the system to be 
an eclipsing binary (EB). Such effects should not be observed if the detected signal
is from a planetary transit. We fitted a \citet{2002ApJ...580L.171M} model to
the odd- and even-numbered transits separately. If the transit depths differed by
more than 3$\sigma$, the system was flagged as a likely false positive. 

We also searched for any evidence of a secondary eclipse. First we fitted the observed
transits with a \citet{2002ApJ...580L.171M} model. The fit was used as a
template for the secondary eclipse. We allowed the eclipse depth and time of opposition
to float freely; all the other relevant parameters were held fixed based on the transit model. If a
secondary eclipse was detected with more than 3$\sigma$ significance,
we then calculated the geometric albedo implied by the depth of secondary
eclipse. If the implied albedo was much larger than 1, we concluded the eclipsing object
is likely to be too luminous to be a planet.
Typically, in each of the {\it K2} Campaigns 5, 6, 7, 8, and 10, approximately $5-10$ 
M-dwarf planetary candidates survived this initial vetting process.

\section{Observations and Data Reductions\label{s:obs}}\label{s:obs}

We here report the follow-up observations 
for the planet candidates around M dwarfs detected by our pipeline. 
The complete list of our candidates will be presented elsewhere
(Livingston et al.\ and other papers in preparation). 
We attempted follow-up observations 
for as many M-dwarf planet hosts as possible. Our selection of targets
included all planet candidates that had not already been validated (to our knowledge), 
with a preference for northern-hemisphere targets for which
our follow-up resources are best suited.
Specifically, we report on the candidates around
\taraa, \tarab, \tarad, \tarac, \tarba, \tarca, \tarcb, \tarcc, \tarcd, \tarce, \tarcf, \tarda, \tardb, and \tardc,
for which we conducted both high-resolution imaging and optical spectroscopy.
This list of M dwarfs covers about half of all candidate planet-hosts 
in the {\it K2} Campaign fields 5, 8, and 10. 
Campaign fields 6 and 7 are located in the southern hemisphere where
our telescope resources are limited. 
The M-dwarf systems we did not follow up are generally fainter objects ($V>15$) 
for which follow-up observations are difficult and time-consuming.

\subsection{Low Dispersion Optical Spectroscopy\label{s:cafos}}\label{s:cafos}

We conducted low dispersion optical spectroscopy with the Calar Alto Faint Object Spectrograph
(CAFOS) on the 2.2~m telescope at the Calar Alto observatory. We observed planet-host candidates
in {\it K2} campaign fields 5 and 8 
(\taraa, \tarab, \tarac, \tarca, \tarcb, \tarcd, \tarce, \tarcf) 
on UT 2016 October 28 and 29, and three stars in field 10 (\tarda, \tardb, \tardc) 
on UT 2017 February 21\footnote{As we describe in Section \ref{s:fitK2}, \tarcc\ (EPIC 220194974) 
turns out to be the planet host, although at first we misidentified \tarcb\ to be the host
of transiting planets and obtained the optical spectrum for \tarcb\ with CAFOS.}. 
Following \citet{2015A&A...577A.128A}, we employed the grism 
``G-100" setup, covering $\sim 4200-8300~\mathrm{\AA}$ with a spectral resolution of 
$R\sim 1500$. The exposure times ranged from 600 s to 2400 s depending on the
magnitude of each star. For long exposures ($>600$ s), we split the exposures into
several small ones so that we can minimize the impact of cosmic rays in the data
reduction. For the absolute flux calibration, we observed Feige 34 as a flux standard
on each observing night. 
We did not observe \tarba\ because this target never rises above 25$^\circ$ elevation at Calar Alto.

\begin{figure}
\centering
\includegraphics[width=8.5cm]{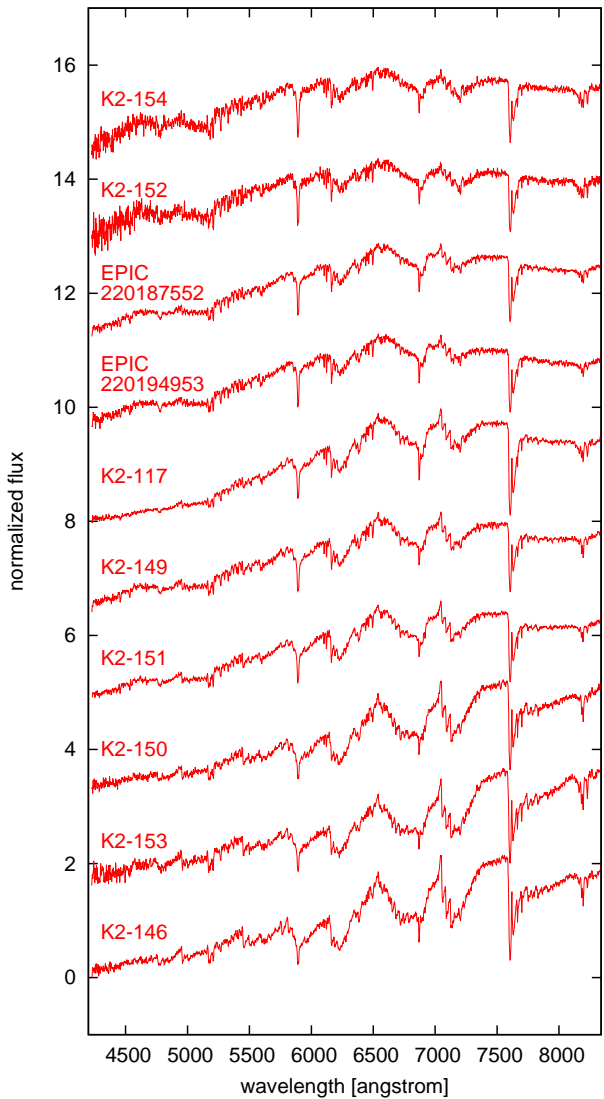}
\caption{Wavelengh-calibrated, normalized optical spectra observed by CAFOS. 
Later M dwarfs are plotted towards the bottom. }
\label{fig:cafos}
\end{figure}
We reduced the data taken by CAFOS in a standard manner using IRAF packages; 
bias subtraction, flat-fielding, sky-subtraction, and extraction of one-dimensional (1D) spectra. 
Wavelength was calibrated using the revised line list of the comparison lamp (Hg-Cd-Ar) 
spectrum \citep{2015A&A...577A.128A}. Finally, we corrected the instrumental response
and converted the flux counts into the absolute fluxes using the extracted 1D spectrum of Feige 34. 
The data for one of the targets, \tarac, were not useful because the signal-to-noise ratio (SNR) of
  the spectrum turned out to be too low.
Figure \ref{fig:cafos} plots the reduced, normalized spectra observed by CAFOS.

\subsection{High Dispersion Spectroscopy\label{s:hires}}\label{s:hires}

In order to estimate stellar physical parameters and check binarity, we obtained high resolution 
optical spectra with various spectrographs. 
\taraa, \tarab, \tarac, \tarba, \tarca, \tarcb, \tarcc, \tarcd, \tarce, \tarcf, and \tardb\ were
observed by High Dispersion Spectrograph \citep[HDS;][]{2002PASJ...54..855N} on the Subaru 8.2~m telescope
between 2015 fall and 2017 summer. For all HDS targets except \tarab, we adopted the standard ``I2a" setup and 
Image Slicer \#2 \citep{2012PASJ...64...77T}, covering the spectral region of $\sim4900-7600\mathrm{\AA}$
with a resolving power of $R\sim 80000$. To avoid a telescope auto-guiding error, 
we adopted the normal slit with its width being $0\farcs6$ ($R\sim 60000$) for \tarab, which
is the faintest in the optical among our targets.

For \tarac, \tarca, \tarcd, \tarce, and \tarcf, we also conducted multi-epoch observations, 
spanning at least a few days, mainly to check the absence of large RV variations 
($\gtrsim 1$ km s$^{-1}$) caused by stellar companions (i.e., EB scenarios). 
Except \tarce, the multi-epoch spectra were taken with the iodine (I$_2$) cell; 
the stellar light, transmitted through the cell,
is imprinted with the iodine absorption lines which are used for the
simultaneous precise calibration of wavelength \citep[e.g.,][]{1996PASP..108..500B}. 
By using the I$_2$ cell, we can improve the RV precision by more than tenfold, 
and can not only rule out the EB scenario but also put a constraint on 
planetary masses, provided that the spectra are obtained at appropriate orbital phases. 
The only drawback is that we need to take one additional I$_2-$free spectrum 
as a template in the RV analysis for each target.

Two-dimensional (2D) HDS data in echelle format were reduced in the standard manner, 
including flat-fielding, scattered-light subtraction, and extraction of 1D spectra for
multiple orders. Wavelength was calibrated based on the Th-Ar emission lamp 
spectra obtained at the beginning and end of each observing night. 
Typical SNR's of the resulting 1D spectra were $\sim 20-50$ per pixel
around sodium D lines.

\begin{figure}
\centering
\includegraphics[width=8.1cm]{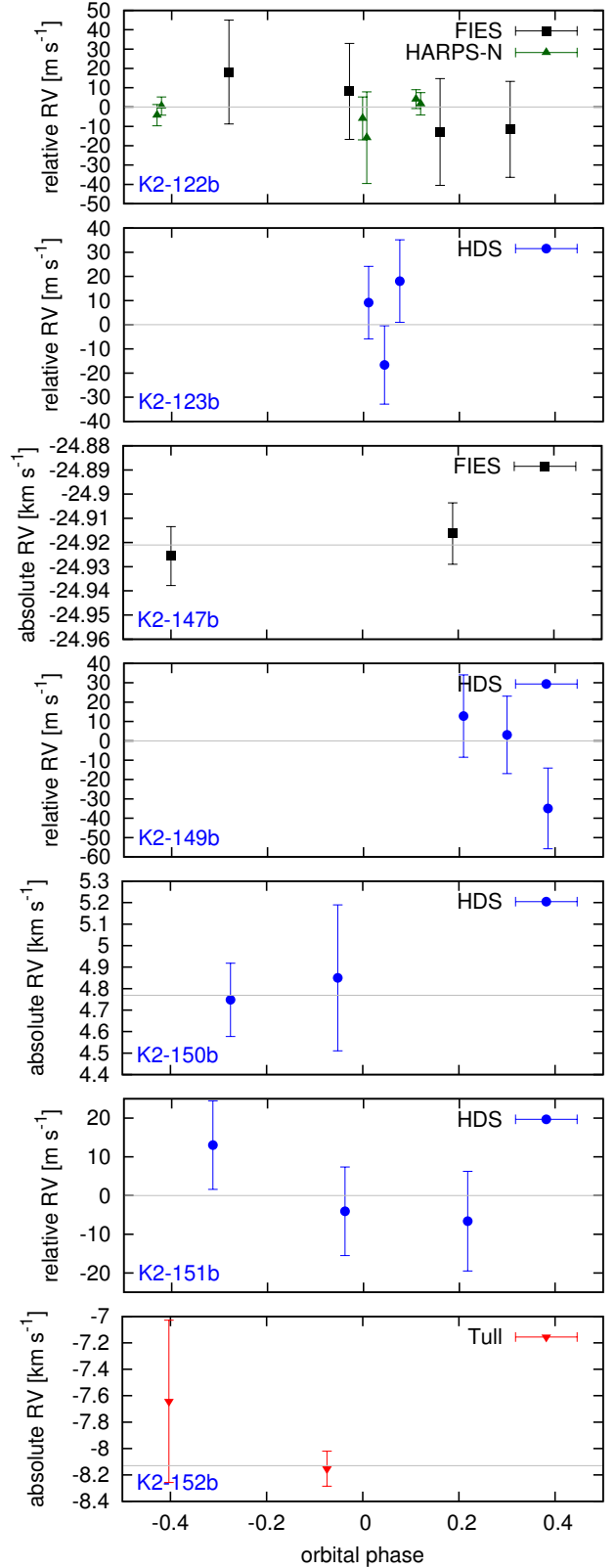}
\caption{RV values folded by the orbital period of each transiting planet. 
Relative RV values are plotted for \tarad, \tarac, \tarcd, and \tarcf, while 
absolute RV values are shown for \tarba, \tarce\ and \tarda. 
Note that for \tarad, the systemic velocity was subtracted from each dataset
to take into account the small RV offset between the FIES and HARPS-N datasets. 
}
\label{fig:rv}
\end{figure}

\begin{table}[tb]
\begin{center}
\small
\caption{Results of RV Measurements}\label{tab:rv}
\begin{tabular}{lcccc}
\hline\hline
$\mathrm{BJD}_\mathrm{TDB}$ & RV & RV error & RV Type & Instrument \\
($-2450000.0$) & (km s$^{-1}$) & (km s$^{-1}$) & & \\\hline
\multicolumn{2}{l}{\bf \tarad} &&& \\
7343.722376 & $-14.6049$ & 0.0248 & absolute & FIES \\
7395.510251 & $-14.6245$ & 0.0248 & absolute & FIES \\
7398.646686 & $-14.5949$ & 0.0269 & absolute & FIES \\
7399.624305 & $-14.6259$ & 0.0276 & absolute & FIES \\
7370.661943 & $-14.3411$ & 0.0049 & absolute & HARPS-N \\
7370.683403 & $-14.3435$ & 0.0058 & absolute & HARPS-N \\
7372.633972 & $-14.3511$ & 0.0111 & absolute & HARPS-N \\
7372.653348 & $-14.3610$ & 0.0237 & absolute & HARPS-N \\
7400.532625 & $-14.3494$ & 0.0055 & absolute & HARPS-N \\
7400.553493 & $-14.3447$ & 0.0047 & absolute & HARPS-N \\
\multicolumn{2}{l}{\bf \tarac} &&& \\
7674.087730 & $0.0156$ & 0.0150 & relative & HDS \\
7675.115382 & $-0.0102$ & 0.0162 & relative & HDS \\
7676.095845 & $0.0245$ & 0.0171 & relative & HDS \\
\multicolumn{2}{l}{\bf \tarba} &&& \\
7893.706393 & $-24.9163$ & 0.0127 & absolute & FIES \\
7931.617000 & $-24.9256$ & 0.0122 & absolute & FIES \\
\multicolumn{2}{l}{\bf \tarcd} &&& \\
7674.002138 & $0.0132$ & 0.0213 & relative & HDS \\
7675.030047 & $0.0034$ & 0.0200 & relative & HDS \\
7675.998989 & $-0.0346$ & 0.0209 & relative & HDS \\
\multicolumn{2}{l}{\bf \tarce} &&& \\
7675.072056 & $4.748$ & 0.171 & absolute & HDS\\
7921.089719 & $4.850$ & 0.339 & absolute & HDS\\
\multicolumn{2}{l}{\bf \tarcf} &&& \\
57674.03764 & $0.0089$ & 0.0115 & relative & HDS \\
7675.094883 & $-0.0082$ & 0.0114 & relative & HDS \\
7676.077393 & $-0.0107$ & 0.0129 & relative & HDS \\
\multicolumn{2}{l}{\bf \tarda} &&& \\
7834.755773 & $-8.153$ & 0.133 & absolute & Tull\\
7954.629452 & $-7.643$ & 0.614 & absolute & Tull\\
\hline
\end{tabular}
\end{center}
\end{table}
For RV targets observed with the I$_2$ cell (\tarac, \tarca, \tarcd, and \tarcf), we
put the reduced 1D spectra into the RV analysis pipeline developed by \citet{2002PASJ...54..873S}
and extracted relative RV values with respect to the I$_2$-out template spectrum 
for each target. Among the four targets, the RV fit did not converge for \tarca, 
which turns out to be a spectroscopic binary (see Sections \ref{s:ircs} and \ref{s:parameter}). 
The results of RV measurements are summarized in Table \ref{tab:rv}. 
Figure \ref{fig:rv} plots the relative RV variation as a function of orbital
phase of each planet candidate; the absence of significant RV variations, along
with the typical RV precision of $10-20$ m s$^{-1}$ for I$_2-$in spectra, completely 
rules out the presence of stellar companions in close-in orbits.

We performed the RV follow-up observations of \tarad\ and \tarba\ using the FIbre-fed \'Echelle 
Spectrograph \citep[FIES;][]{1999anot.conf...71F, 2014AN....335...41T} mounted at the 2.56~m Nordic 
Optical Telescope (NOT) of Roque de los Muchachos Observatory (La Palma, Spain). 
We collected 4 high-resolution spectra ($R\sim 67,000$) of \tarad\ between November 2015 and January 2016, 
and 2 intermediate-resolution spectra ($R\sim 47,000$) of \tarba\ in May and June 2017, as part of the 
observing programs P52-201 (CAT), P52-108 (OPTICON), and P55-019.
Three consecutive exposures of 900-1200 s were secured to remove cosmic ray hits, 
leading to an SNR of 25-30 per pixel at 5800\,\AA. We followed the observing strategy 
described in \citet{2010ApJ...720.1118B} and \citet{2013A&A...557A..74G}, and traced 
the RV intra-exposure drift of the instrument by acquiring long-exposed (T$_\mathrm{exp}$\,=\,35 s) 
Th-Ar spectra immediately before and after each observation. The data reduction was 
performed using standard IRAF and IDL routines, which include bias subtraction, flat fielding, 
order tracing and extraction, and wavelength calibration. 
The RVs were determined by multi-order cross-correlation against a spectrum of the M2-dwarf GJ\,411 
that was observed with the same instrumental set-ups as the two target stars, and for which we adopted 
an absolute RV of $-84.689$ km~s$^{-1}$.

We also acquired 6 high-resolution spectra ($R\sim 115,000$) of \tarad\
using the HARPS-N spectrograph \citep{2012SPIE.8446E..1VC} mounted at the 3.58~m Telescopio Nazionale Galileo (TNG) of Roque de los Muchachos Observatory (La Palma, Spain). Two consecutive exposures of 1800 s were acquired at 3 different epochs between December 2016 and January 2017, 
as part of the CAT and OPTICON programs CAT15B\_35 and OPT15B\_64, 
using the second HARPS-N fiber to monitor the sky background. Unfortunately, the spectra 
taken on $\mathrm{BJD}=2457372$ are affected by poor sky conditions. We reduced the data 
using the dedicated off-line pipeline. The SNR is between 5 and 20 per pixel at 5800\,\AA. 
RVs were extracted by cross-correlating the extracted echelle spectra with the M2 numerical mask 
(Table \ref{tab:rv}).

We observed \tarda\ and \tardc\ with the Harlan J. Smith 2.7~m telescope and its Tull Coud\'e high-resolution 
($R=60,000$) optical spectrograph \citep{1995PASP..107..251T} at McDonald Observatory. We obtained one reconnaissance spectrum of \tarda\ in March 2017 and a second one in July 2017. 
We also collected one spectrum of \tardc\ in March 2017. Exposure times ranged from 29 to 50 minutes, 
due to the faintness of these stars in the optical.
The spectra were all bias-subtracted, flat-field divided and extracted using standard IRAF 
routines. For the wavelength calibration, we use Th-Ar calibration exposures taken adjacent to the science observations.    
We analyzed the spectra using our {\it Kea} code \citep{2016PASP..128i4502E} to determine stellar parameters. 
{\it Kea} is not well suited to derive accurate parameters for
cooler stars, but the results showed that both stars are cool ($T_{\rm eff}\sim 4000$\,K) main sequence stars.   
In Section \ref{s:specmatch}, we will perform a more uniform analysis to estimate stellar parameters.

\subsection{High Contrast Imaging\label{s:ircs}}\label{s:ircs}

In transit surveys, typical false positives arise from background or hierarchical-triple EBs. 
High resolution imaging is especially useful 
to constrain background EB scenarios, and thus has intensively been used for planet validations
\citep[e.g.,][]{2017AJ....154..207D}. 
To search for nearby companions that could be could be the source of the observed transit-like signal, we conducted
high resolution imaging using the adaptive-optics system \citep[AO188;][]{2010SPIE.7736E..0NH} 
with the High Contrast Instrument \citep[HiCIAO;][]{2010SPIE.7735E..30S}
for \tarab\ and \tarad\ and the Infrared Camera and Spectrograph \citep[IRCS;][]{2000SPIE.4008.1056K} 
for the other systems, both mounted on the Subaru telescope between 2015 winter and 2017 summer.

For the HiCIAO observation, we adopted the same observing scheme as described in 
\citet{2016ApJ...825...53H}, except that we employed the angular differential imaging
\citep[ADI;][]{2006ApJ...641..556M} for \tarab. 
With the three-point dithering and $H-$band filter, 
a total of 11 unsaturated frames after co-addition were obtained with AO for \tarab, 
resulting in the total exposure time of 1135 s. 
For \tarad, we obtained three saturated frames (after co-addition) with two-point dithering, 
corresponding to the total exposure time of 450 s. We also took two unsaturated frames 
for absolute flux calibration using a neutral-density filter.

HiCIAO data were reduced with the ACORNS pipeline developed by \citet{2013ApJ...764..183B}
for the removal of biases and correlated noises, hot pixel masking, flat-fielding, and distortion correction. 
We then aligned and median-combined the processed frames to obtain the highest contrast image.
The resulting full width at half maximum (FWHM) of the combined images were $\sim 0\farcs07$. 
We visually inspected the combined images for \tarab\ and \tarad, and found two neighboring faint 
companions to the northwest of \tarab. 
The brighter of the two is located $9\farcs1$ away from \tarab\ with $\Delta m_H=6.7$ mag, while
the fainter is $8\farcs7$ away from \tarab\ with $\Delta m_H=7.7$ mag. 
Checking the SDSS catalog \citep{2012ApJS..203...21A}, we identified a star around the coordinate 
where two faint stars were detected, and found its relative magnitude to be $\Delta m_r=6.4$ mag. 
These faint stars are inside the photometric aperture for the {\it K2} light curve, 
but the optical and near infrared magnitudes imply that these cannot produce the deep transit 
signal detected for \tarab. We detected no nearby companion in the combined image
of \tarad.

Regarding IRCS observations, we conducted AO imaging using each target itself as the natural guide 
for AO with the $H-$band filter. Adopting the fine sampling mode ($1~\mathrm{pix}=0\farcs02057$)
and five-point dithering, we ran two kinds of sequences for
each target. The first sequence consists of long exposures to obtain saturated frames of the
targets, which are used to search for faint nearby companions. 
The total exposure time varied widely for each target, but typically $\sim 360$ s for
a $m_H=10$ mag star. The saturation radii were less than $0\farcs05$ for all frames. 
As the second sequence, we also took unsaturated frames with much shorter exposures, 
and used these frames for absolute flux calibrations. 

\begin{figure*}
\centering
\includegraphics[width=18cm]{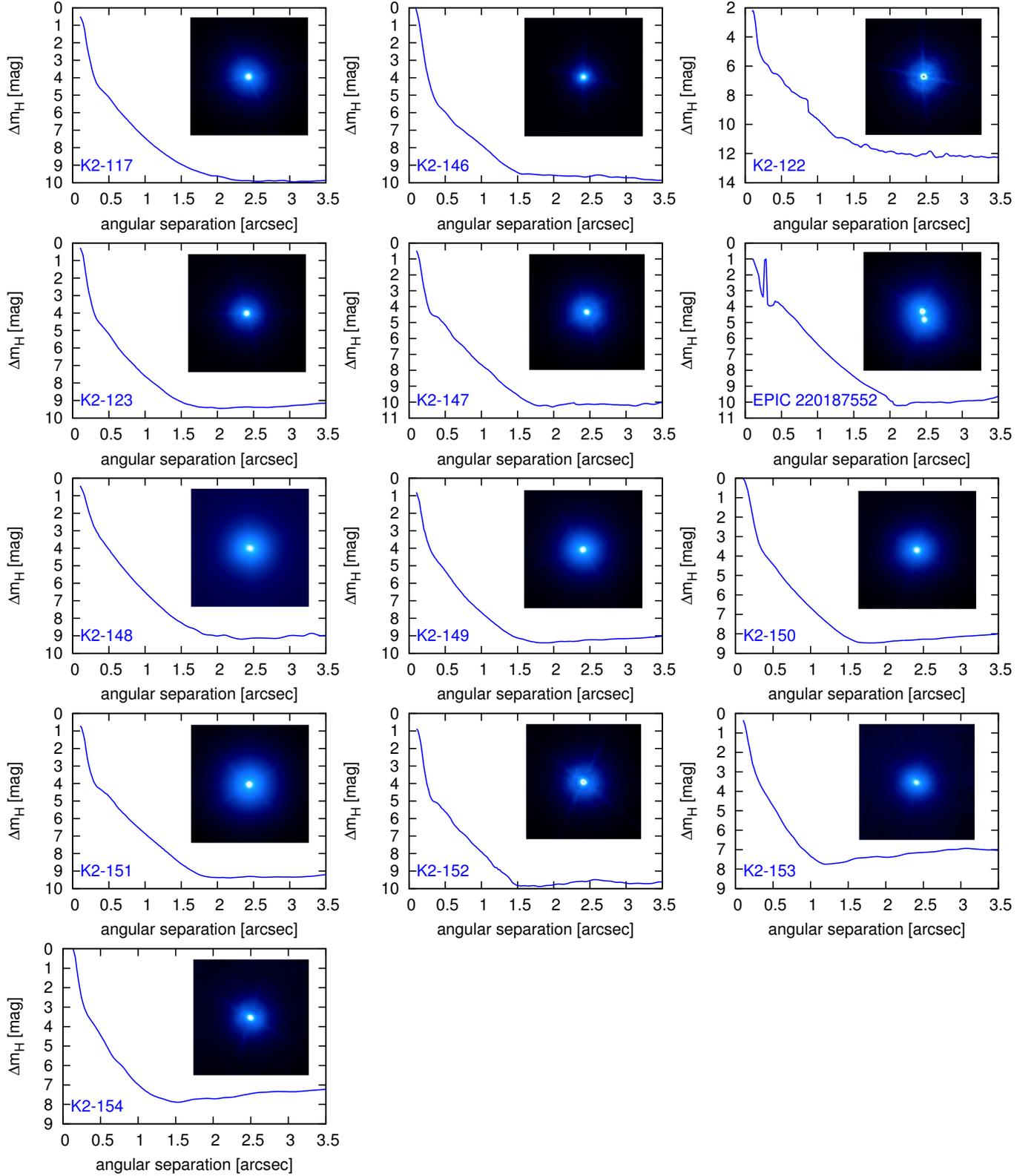}
\caption{$5\sigma$ contrast curves in the $H$ band as a function of angular separation from
the centroid for {\it K2} planet-host candidates. The insets display the saturated combined images
with FoV of $4^{\prime\prime}\times4^{\prime\prime}$. \tarca\ is clearly a multiple-star system, 
and we conclude that the candidate is a false positive. }
\label{fig:contrast}
\end{figure*}

Following \citet{2016ApJ...820...41H}, we reduced the raw IRCS data:
subtraction of the dark current, flat fielding, and distortion correction, before
aligning and median-combining the frames for each target. 
The combined images were respectively generated for saturated and unsaturated frames. 
We visually checked the combined saturated image for each target, 
in which the field-of-view (FoV) is $\sim 16^{\prime\prime}\times16^{\prime\prime}$. 
Most importantly, we found that \tarca\ consists of two stars of similar magnitude 
separated by $\sim 0\farcs3$ from each other (Figure \ref{fig:contrast}). 
In the same image, we also found a faint star at $\sim 6^{\prime\prime}$ away 
from \tarca\ with $\Delta m_H\sim 8$ mag. 
\tarcb\ and \tarcc\ were both imaged in the same combined frame. 
\tarba's combined image also exhibits a possible faint star ($\Delta m_H\sim 9.5$ mag) 
in the south, but with a low SNR, separated by $4\farcs6$. 
We found no bright nearby stars in the FoV for the other targets.

To estimate the detection limit of faint nearby sources in the combined images, 
we drew $5\sigma$ contrast curve for each object. 
To do so, we first convolved the saturated images with each convolution radius being
half of FWHM. We then calculated the scatter of 
the flux counts in the narrow anulus as a function of angular separation from the 
target's centroid. Finally, we obtained the target's absolute flux by aperture photometry
using the unsaturated frames for each target with aperture diameter being FWHM, 
and normalized the flux scatter in the anulus by dividing by the photometric value
after adjusting the exposure times for saturated and unsaturated combined images. 
Figure \ref{fig:contrast} displays the $5\sigma$ contrast curves for all objects,
along with the $4^{\prime\prime}\times4^{\prime\prime}$ combined images of the targets 
in the insets. Note that as we show in Section \ref{s:fitK2}, \tarcb\ and \tarcc\
are imaged in the same frame, but since \tarcc\ is likely the host of transiting 
planets, we show the contract curve around it.

\subsection{Follow-up Transit Observations}\label{s:followup}

\subsubsection{OAO 188cm/MuSCAT}

On 2016 September 20, we conducted a photometric follow-up observation of a transit of \tarcf b 
with the Multi-color Simultaneous Camera for studying Atmospheres of Transiting exoplanets 
\citep[MuSCAT;][]{2015JATIS...1d5001N} on the 1.88~m telescope at Okayama Astronomical Observatory (OAO). 
MuSCAT is equipped with three 1k$\times$1k  CCDs with a pixel scale of 0\farcs36~pixel$^{-1}$, 
enabling us to obtain three-band images simultaneously through the SDSS 2nd-generation 
$g'$, $r'$, and $z_\mathrm{s}$-band filters. We set the exposure times to 60, 10, and 25~s for the 
$g'$, $r'$, and $z_\mathrm{s}$ bands, respectively.
We observed the target star along with several bright comparison stars for $\sim$3.8~h, which covered well
the expected $\sim$1.5-h duration transit. The sky was photometric except for $\sim$0.9~h near the 
end of the observation, when clouds passed; we omit the data during this period from the subsequent 
data reduction process. As a result, 166, 749, and 354 images were obtained in the $g'$, $r'$, and $
z_\mathrm{s}$ bands, respectively, through clear skies.

The observed images were dark-subtracted, flat-fielded, and corrected for non-linearlity of each detector. 
Aperture photometry was performed with a customized pipeline \citep{2011PASJ...63..287F} for the target 
star and three similar-brightness stars for comparison, one of which, however, was saturated on the $g'$-band 
images and omitted from the rest of the analysis for this band. The aperture radius for each band was optimized 
so that the apparent dispersion of a relative light curve (a light curve of the target star divided by that of the 
comparison stars) was minimized. As a result, the radii of 11, 13, and 12 pixels were adopted for 
the $g'$, $r'$, and $z_\mathrm{s}$ bands, respectively.

\subsubsection{IRSF 1.4 m/SIRIUS}\label{sec:irsf_obs}

On 2016 October 5 UT, we also conducted a follow-up transit observation with the Simultaneous Infrared Imager 
for Unbiased Survey \citep[SIRIUS;][]{2003SPIE.4841..459N} on the IRSF 1.4~m telescope at South African 
Astronomical Observatory. SIRIUS is equipped with three 1k$\times$1k HgCdTe detectors with the pixel scale of 
0\farcs45~pixel$^{-1}$, enabling us to take three near-infrared images in $J$, $H$, and $K_\mathrm{s}$ bands simultaneously. 
Setting the exposure times to 30~s with the dead time of about 8~s for all bands, we continued the observations 
for $\sim$2.4~h covering the expected transit time. As a result, 232 frames were obtained in each band.

The observed frames were analyzed in the same manner as the MuSCAT data. For the flat-fielding, we used 14, 
14, and 36 twilight sky frames taken on the observing night for the $J$-, $H$-, and $K_\mathrm{s}$-band data, respectively. 
We applied aperture photometry for the target and two comparison stars for all bands. However, we found that the 
brighter comparison star was saturated in the $H-$band data and was thus useless. With only the fainter comparison star, we 
could not achieve a sufficiently high photometric precision to extract the transit signal, and therefore we 
decided to ignore the $H$-band data from the subsequent analyses. We selected 9~pixels as the optimal 
aperture radii for both $J$ and $K_\mathrm{s}$ band data.

\section{Data Analyses and Validation of Planet Candidates\label{s:analysis}}\label{s:analysis}

\subsection{Estimation of Spectroscopic Parameters\label{s:parameter}}\label{s:parameter}
\subsubsection{Spectral Types}

Based on the low resolution spectra obtained by CAFOS, we measured the spectral types (SpT) 
for the target stars. Following \citet{2015A&A...577A.128A}, we measured a suite of (31) 
spectral indices for each CAFOS spectrum. 
\citet{2015A&A...577A.128A} found that five indices (TiO 2, TiO 5, PC1, VO-7912, and 
Color-M) amongst all have the best correlations with SpT and thus we converted each of the measured 
five indices listed in Table \ref{hyo0} into SpT through the polynomials given by \citet{2015A&A...577A.128A}, 
with revised coefficients \citep{2015PhDT........97A}. 
We then took the weighted mean of the calculated SpT values to obtain the final value for
each target and round those mean spectral types to the nearest standard subtypes
(e.g., M0.0, M0.5, M1.0, $\cdots$), which are listed in Table \ref{hyo1}. 
The scatter of the calculated SpT values from the five indices for each object is generally
less than 0.5 subtype, which is comparable to the fiducial measurement error in SpT 
by the present method. 
The converted SpT values for \taraa\ have a relatively large scatter (standard deviation $=0.523$ 
subtype), which might be due to passage of clouds or other bad weather conditions.

\begin{table}[tb]
\begin{center}
\small
\caption{Spectral Indices by CAFOS Spectroscopy. 
}\label{hyo0}
\begin{tabular}{lccccc}
\hline\hline
Star & TiO 2 & TiO 5 & PC1 & VO-7912 & Color-M \\\hline
\taraa & 0.826 & 0.662 & 1.037 & 0.998 & 0.752 \\ 
\tarab & 0.641 & 0.423 & 1.157 & 1.072 & 1.045 \\ 
\tarac & 1.061 & 0.998 & 0.935 & 0.980 & 0.556 \\ 
{\scriptsize \tarca} & 0.866 & 0.730 & 0.984 & 0.999 & 0.733 \\ 
{\scriptsize \tarcb} & 0.877 & 0.742 & 0.978 & 0.994 & 0.713 \\ 
\tarcd & 0.807 & 0.635 & 1.012 & 1.004 & 0.778 \\ 
\tarce & 0.697 & 0.481 & 1.137 & 1.049 & 1.057 \\ 
\tarcf & 0.789 & 0.622 & 1.023 & 1.010 & 0.816 \\ 
\tarda & 0.919 & 0.775 & 0.949 & 0.998 & 0.748 \\ 
\tardb & 0.662 & 0.472 & 1.163 & 1.073 & 1.269 \\ 
\tardc & 0.888 & 0.748 & 0.955 & 0.995 & 0.753 \\ 
\hline
\end{tabular}
\end{center}
\end{table}

We also checked if the target stars are dwarf stars and not M giants, by inspecting
the index ``Ratio C" \citep{1991ApJS...77..417K}, which is a good indicator of surface gravity. 
As described in \citet{2015A&A...577A.128A}, stars with a low surface gravity should have
a value of Ratio C lower than $\sim 1.07$, but all the targets listed in Table \ref{hyo1}
show higher Ratio C values, by which we safely conclude that
those stars observed by CAFOS are all M dwarfs.

\subsubsection{Atmospheric and Physical Parameters\label{s:specmatch}}\label{s:specmatch}

In order to estimate the precise atmospheric and physical parameters of the target stars, 
we analyzed high resolution optical spectra obtained in Section \ref{s:hires}. 
We made use of \texttt{SpecMatch-Emp} developed by \citet{2017ApJ...836...77Y}. 
\texttt{SpecMatch-Emp} uses a library of optical high resolution spectra 
for hundreds of well-characterized FGKM stars collected by the California Planet Search; 
it matches an observed spectrum of unknown propety to library stars, by which the best-matched 
spectra and their stellar parameters (the effective temperature $T_\mathrm{eff}$, stellar 
radius $R_s$, and metallicity [Fe/H]) are found for the input spectrum while the RV shift 
and rotation plus instrumental line-broadening are simultaneously optimized. 
\texttt{SpecMatch-Emp} is particularly useful for late-type stars, for which 
spectral fitting using theoretical models often has large systematics due to imperfection 
of the molecular line list in the visible region.

Since \texttt{SpecMatch-Emp} is developed for optical spectra obtained by 
Keck/HIRES, we converted our spectra taken by Subaru/HDS, etc, into the same
format as HIRES. To check the validity of applying \texttt{SpecMatch-Emp} 
to those spectra taken by other instruments, for which spectral resolutions and
pixel-samplings are slightly different from those of HIRES, we put several spectra 
collected by Subaru/HDS in the past campaigns \citep[e.g.,][]{2014ApJ...783....9H}
into \texttt{SpecMatch-Emp} and compared the 
outputs with literature values. Consequently, we found that the output $T_\mathrm{eff}$,
$R_s$, and [Fe/H] are all consistent with the literature values within $2\sigma$ (typically 
within $1\sigma$), and we justified the validity of applying \texttt{SpecMatch-Emp} 
to our new spectra.

Inputting our high resolution spectra to \texttt{SpecMatch-Emp}, 
we obtained the stellar spectroscopic parameters. We discarded \tarca\ from this analysis, since
\tarca\ was found to be a double (in fact triple) star revealed by the AO imaging (Section \ref{s:ircs}). 
The output parameters ($T_\mathrm{eff}$, $R_s$, and [Fe/H]) are listed in Table \ref{hyo1}. 
To estimate the other stellar parameters (i.e., stellar mass $M_s$, surface gravity $\log g$, 
and luminosity $L_s$), we adopted the empirical formulas derived by \citet{2015ApJ...804...64M}, 
who gave empirical relations of stellar mass and radius as a function of the absolute
$K_s-$band magnitude and [Fe/H]. Assuming that \texttt{SpecMatch-Emp}'s output parameters
follow independent Gaussians with their $\sigma$ being the errors returned by 
\texttt{SpecMatch-Emp}, we performed Monte Carlo simulations and converted
$T_\mathrm{eff}$, $R_s$, and [Fe/H] into $M_s$, $\log g$, and $L_s$ through
the absolute $K_s-$band magnitude. Those estimates are also summarized in 
Table \ref{hyo1}. In the same table, we also list the distance $d$ calculated from the apparent
and absolute $K_s-$band magnitudes.

\begin{table*}[tb]
\begin{center}
\small
\caption{Stellar Parameters by Optical Low and High Resolution Spectroscopy. 
}\label{hyo1}
\begin{tabular}{lccccccccc}
\hline\hline
EPIC ID & {\it K2} ID & SpT & $T_\mathrm{eff}$ (K) & [Fe/H] (dex) & $R_s$ ($M_\odot$) & $M_s$ ($M_\odot$) & $\log g$ (dex) & $L_s$ ($L_\odot$) & $d$ (pc)\\\hline
211331236 & K2-117 & M$1.0$V & $3676\pm70$ & $-0.22\pm 0.12$ & $0.513\pm 0.051$ & $0.532\pm 0.056$ & $4.747\pm 0.046$ & $0.044\pm 0.009$ & $100\pm 14$ \\
211924657 & K2-146 & M$3.0$V & $3385\pm70$ & $-0.02\pm 0.12$ & $0.350\pm 0.035$ & $0.358\pm 0.042$ & $4.906\pm 0.041$ & $0.015\pm 0.003$ & $86\pm 11$ \\
212006344 & K2-122 & $-$ & $3903\pm70$ & $0.37\pm 0.12$ & $0.612\pm 0.061$ & $0.644\pm 0.061$ & $4.677\pm 0.051$ & $0.079\pm 0.017$ & $74\pm 11$ \\
212069861 & K2-123 & $-$ & $3880\pm70$ & $-0.02\pm 0.12$ & $0.592\pm 0.059$ & $0.615\pm 0.060$ & $4.686\pm 0.049$ & $0.072\pm 0.016$ & $156\pm 24$ \\
213715787 & K2-147 & $-$ & $3672\pm70$ & $0.19\pm 0.12$ & $0.554\pm 0.055$ & $0.583\pm 0.059$ & $4.720\pm 0.048$ & $0.051\pm 0.011$ & $88\pm 13$ \\
220187552 & $-$ & M$0.5$V & $-$ & $-$ & $-$ & $-$ & $-$ & $-$ & $-$\\
220194953 & $-$ & M$0.5$V & $3854\pm70$ & $-0.04\pm 0.12$ & $0.575\pm 0.058$ & $0.598\pm 0.059$ & $4.699\pm 0.049$ & $0.066\pm 0.014$ & $121\pm 18$ \\
220194974 & K2-148 & $-$ & $4079\pm70$ & $-0.11\pm 0.12$ & $0.632\pm 0.063$ & $0.650\pm 0.061$ & $4.653\pm 0.051$ & $0.101\pm 0.022$ & $121\pm 19$ \\
220522664 & K2-149 & M$1.0$V & $3745\pm70$ & $0.11\pm 0.12$ & $0.568\pm 0.057$ & $0.595\pm 0.059$ & $4.707\pm 0.048$ & $0.049\pm 0.011$ & $118\pm 18$ \\
220598331 & K2-150 & M$2.5$V & $3499\pm70$ & $0.09\pm 0.12$ & $0.436\pm 0.044$ & $0.457\pm 0.051$ & $4.822\pm 0.043$ & $0.026\pm 0.006$ & $110\pm 15$ \\
220621087 & K2-151 & M$1.5$V & $3585\pm70$ & $-0.32\pm 0.12$ & $0.429\pm 0.043$ & $0.440\pm 0.050$ & $4.820\pm 0.043$ & $0.028\pm 0.006$ & $62.7\pm 8.8$ \\
201128338 & K2-152 & M$0.0$V & $3940\pm70$ & $0.09\pm 0.12$ & $0.631\pm 0.063$ & $0.654\pm 0.061$ & $4.657\pm 0.051$ & $0.087\pm 0.019$ & $112\pm 18$ \\
201598502 & K2-153 & M$3.0$V & $3720\pm70$ & $-0.26\pm 0.12$ & $0.495\pm 0.050$ & $0.512\pm 0.055$ & $4.761\pm 0.045$ & $0.043\pm 0.009$ & $126\pm 18$ \\
228934525 & K2-154 & M$0.0$V & $3978\pm70$ & $0.19\pm 0.12$ & $0.649\pm 0.065$ & $0.672\pm 0.061$ & $4.645\pm 0.052$ & $0.096\pm 0.021$ & $133\pm 21$ \\
\hline
\end{tabular}
\end{center}
\end{table*}

\subsubsection{Cross-correlation Analysis}

In addition to estimating stellar parameters from the high resolution spectra, 
we also analyzed the line profile for each target. In the case that a transit-like
signal is caused by an eclipsing spectroscopic binary of similar size, 
we expect to see a secondary line or distortion of the profile in the spectra, 
depending on the orbital phase of the binary. 
Using the cross-correlation technique, we computed the averaged spectral line
profiles so that we can check for the presence of line blending. 
In doing so, we cross-correlated each observed spectrum (without the I$_2$ cell) with the 
numerical binary mask \citep[M2 mask; see e.g.,][]{2013A&A...549A.109B} developed 
for the RV analysis of HARPS-like spectrographs. From each observed spectrum, 
we extracted the spectral segments whose wavelengths are covered by the binary mask, 
and cross-correlated each segment with the mask as a function of Doppler shift (RV). 
We then took a weighted average of the cross-correlation profiles to get the normalized 
line profile for each object.

\begin{figure}
\centering
\includegraphics[width=8.5cm]{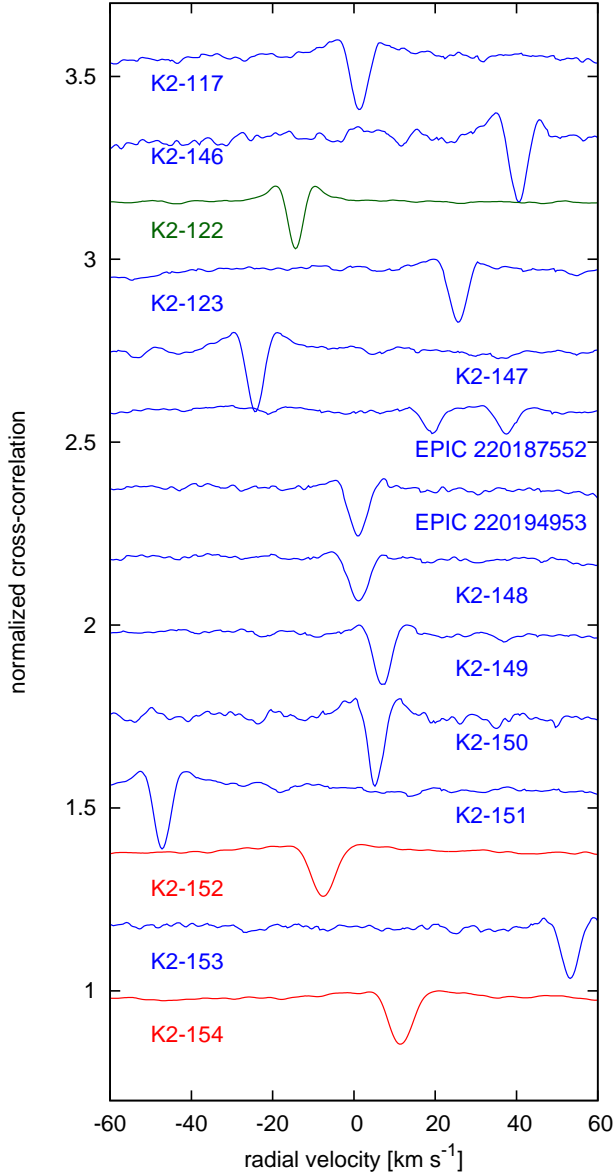}
\caption{Averaged and normalized cross-correlations between the observed spectra and 
M2 binary mask. Cross-correlations based on the HDS, HARPS-N, Tull spectra are shown in blue, 
green, and red, respectively. The Earth's motion is corrected and RV value is given with respect to
the barycenter of the solar system. }
\label{fig:ccf}
\end{figure}
Figure \ref{fig:ccf} displays the line profiles for the observed stars. 
For the targets with multi-epoch observations, we show the cross-correlation 
profiles with the highest SNR. 
Except \tarca, all stars exhibit single-line profiles, though
the cross-correlation continuum looks noisier for particularly cool stars 
(\tarab\ and \tarce), which is most likely due to the more complicated 
molecular absorption features. \tarca\ clearly shows the secondary line in the 
cross-correlation profile, as we expected from Figure \ref{fig:contrast}; due to the small 
angular separation ($\sim 0\farcs3$), the fluxes from the two stars both entered
the spectrograph during our HDS observation. The difference in positions of
the two lines implies that the two stars have a relative Doppler-shift to each other, 
suggesting that either of the two has a stellar companion which is most likely 
responsible for the transit-like signal detected in the {\it K2} light curve. 
Therefore, we concluded that \tarca\ is a hierarchical triple system, in which two stars 
among the three are an EB. We will revisit this system in Section \ref{s:individual}.

From the cross-correlation profile, we also measured the absolute RV for each target. 
Since Subaru/HDS (without the I$_2$ cell) and McDonald 2.7m/Tull are neither stabilized spectrographs 
nor do they obtain simultaneous reference spectra like HARPS/HARPS-N, 
it is difficult to trace the small wavelength drift during a night, which prohibits accurate RV 
measurements. In order to correct for the wavelength drift of each spectrum, 
we extracted the spectral segment including strong telluric absorption lines 
(primarily $6860-6920~\mathrm{\AA}$), and cross-correlated it against a theoretical telluric 
transmission spectrum at the summit of Mauna Kea, generated by using 
line-by-line radiative transfer model \citep[LBLRTM;][]{2005JQSRT..91..233C}. 
Stellar RVs and wavelength drifts are measured by inspecting the peaks (bottoms)
of the cross-correlation profiles for stellar and telluric segments, respectively. 
The final RV values (Table \ref{tab:rv}) are recorded by subtracting the two RV values. 
Note that the resulting wavelength drift is typically less than $0.5$ km s$^{-1}$
(less than half a pixel for HDS). 
Regarding \tarce\ and \tarda, we obtained multiple spectra for absolute RV measurements, 
which are plotted in Figure \ref{fig:rv} as a function of the candidates' phase; 
no significant RV variation is seen for both objects.

\subsection{Light Curve Analysis\label{s:lightcurve}}\label{s;lightcurve}
\subsubsection{Fitting {\it K2} Light Curves\label{s:fitK2}}\label{s:fitK2}

In order to estimate the most precise parameters of each planet candidate, 
we compared the light curves for the same objects produced by three
different pipelines: our own light curves (Section \ref{s:k2reduction}), ones by \citet{2014PASP..126..948V}, 
and ones by EVEREST \citep{2016AJ....152..100L, 2017arXiv170205488L}. 
As a result, we found that for our sample, the EVEREST light curves generally 
provided the best precision in terms of the scatter of the baseline flux.
We thus used EVEREST light curves to estimate the final transit parameters. 
For the three targets in {\it K2} field 10, since EVEREST light curves have not been 
published yet, we employed the light curves by \citet{2014PASP..126..948V}.

We reduced the light curves in the following steps. 
First, using the reduced light curve products, 
we split each target's light curve into segments, each spanning $6-9$ days, and detrended each
segment by fitting with a fifth-order polynomial to get a normalized light curve. Then, 
based on the preliminary ephemerides obtained in Section \ref{s:k2}, we further 
extracted small segments around transit signals, in which the baseline spans
$2.5$ times the duration of the transit towards both sides from the transit center
for each planet candidate. 
These light curve segments around transits were simultaneously fitted for each 
planet candidate.

We fitted all the light curve segments
simultaneously to obtain the global transit parameters as well as check
possible transit timing variations (TTVs). The global transit parameters are
the scaled semi-major axis $a/R_s$, transit impact parameter $b$, 
limb-darkening coefficients $u_1$ and $u_2$ for the quadratic law, and planet-to-star 
radius ratio $R_p/R_s$. We fixed the orbital eccentricity at $e=0$. 
In addition to these, we introduced the parameters
describing the flux baseline, for which we adopted a linear function of time, and
time of the transit center $T_c$ for each transit (segment). 
To take into account the long cadence of {\it K2} observation, we 
integrated the transit model by \citet{2009ApJ...690....1O} over $~29.4$ minutes
to compare the model with observations.

\begin{figure*}
\centering
\includegraphics[width=18.5cm]{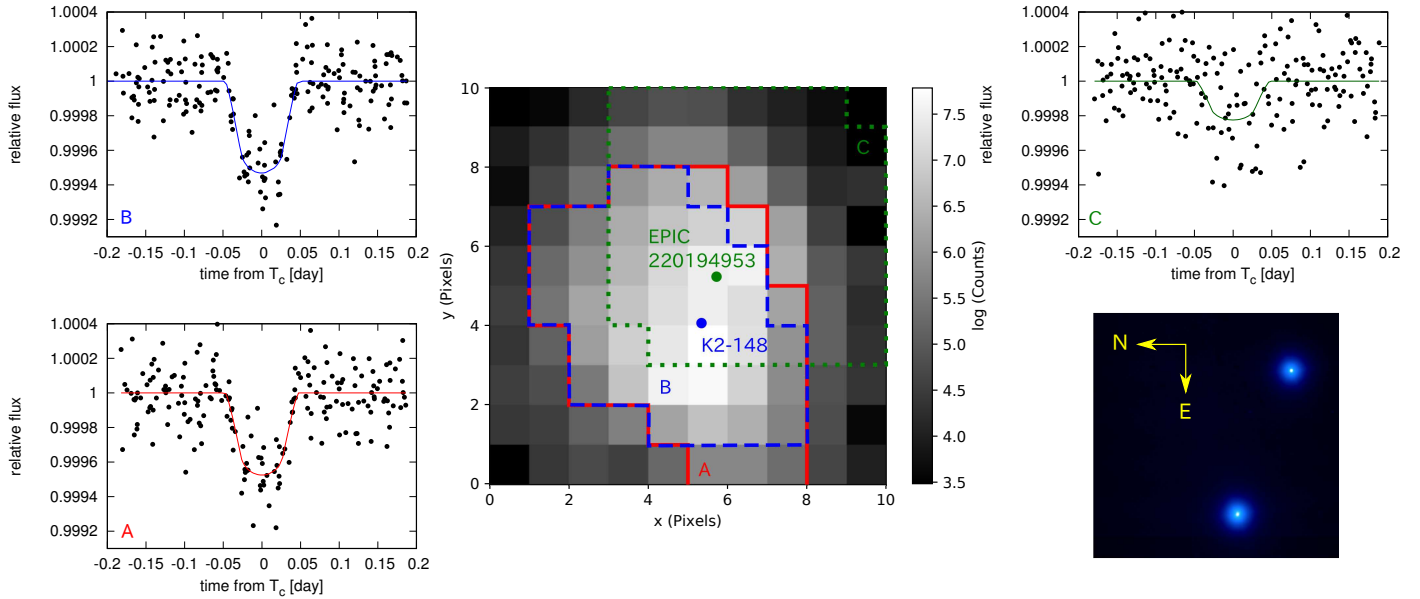}
\caption{EVEREST light curves (left panels and top right panel) produced by different 
apertures (central panel) for \tarcb\ and \tarcc\ (EPIC 220194974). The light curves are folded by the period of
\tarcc c ($=6.92$ days). The right bottom panel shows
a high-resolution image with FoV of $15^{\prime\prime}\times15^{\prime\prime}$ taken by Subaru/IRCS;
the upper right and lower left stars correspond to \tarcb\ and \tarcc, respectively. 
}
\label{fig:aperture}
\end{figure*}
Following \citet{2015ApJ...799....9H}, we first minimized the $\chi^2$
statistic by Powell's conjugate direction method
\citep[e.g.,][]{1992nrca.book.....P} to obtain the best-fit values for
all the parameters, and fixed the baseline parameters for each segment
at these values. We then implemented Markov Chain Monte Carlo (MCMC)
simulations to estimate the posterior distribution of the remaining
fitting parameters. We imposed Gaussian priors on $u_1+u_2$ and
$u_1-u_2$ based on the theoretical values by
\citet{2013A&A...552A..16C}; the central values for $u_1$ and $u_2$
were derived by interpolation for each target using the stellar
parameters listed in Table \ref{hyo1}, and we adopted the dispersion
of Gaussians as 0.1.  At first we assigned an uncertainty to each
  {\it K2} data point equal to the observed scatter in neighboring
  flux values, which sometimes led to a very small or large reduced
  $\chi^2$, presumably due to non-stationary noise.  To obtain
  reasonable uncertainties in the fitted parameter values, we rescaled
  the flux uncertainties such that the reduced $\chi^2$ was equal to unity, before
  performing the MCMC analysis. We adopted the median, and 15.87 and 84.13
percentiles of the marginalized posterior distribution as the central
value and its $\pm 1\sigma$ for each fitting parameter.

\tarcb\ and \tarcc\ are separated by $\sim 9\farcs4$, and the photometric
apertures used to produce EVEREST light curves for those objects involve
at least a part of both stars. In order to identify which of the two stars is the
source of transit signals, we analyzed three different light curves provided by 
EVEREST: the EVEREST version 2.0 light curves for \tarcc\ (EPIC 220194974) (A) and \tarcb\ (B), 
and EVEREST version 1.0 light curve for \tarcb\ (C). The apertures used
to produce the three light curves are shown in the central panel of Figure \ref{fig:aperture}. 
As a result of analyzing and fitting each light curve folded by the period of
\tarcc c, we found that light curves based on
apertures A and B exhibit similar depths in the folded transits, but the one
with aperture C shows a much shallower transit (almost invisible; Figure \ref{fig:aperture}). 
Since a significant fraction of light from \tarcc\ is missing for aperture C, 
\tarcc\ is likely the host of the transiting planet candidates\footnote{We also 
analyzed our own light curves using customized apertures with smaller numbers
of pixels, but the transit signals became invisible owing to the larger scatter in flux.}. 
We thus performed the further analysis below based on this assumption. 
Note that we found a similar trend when the light curve was folded by the period of \tarcc b, 
but with a lower SNR.

To estimate the planetary parameters for \tarcc b to \tarcc d, we need
to know the contamination (dilution) factor from \tarcb\ for the photometric
aperture we adopt. In doing so, we estimated the flux ratio between 
\tarcb\ and \tarcc\ in the Kepler ($Kp$) band by the following procedure
\footnote{The $Kp$ magnitudes are
reported to be 12.856 and 12.975 for \tarcb\ and \tarcc, respectively. However, 
the {\it K2} pixel image ($Kp-$band) and our AO image by IRCS (Figure \ref{fig:aperture}; $H-$band) 
both imply that \tarcc\ is brighter than \tarcb, suggesting \tarcb\ is a later-type star 
the \tarcc\ and the reported $Kp$ magnitudes are inaccurate.}. 
Adopting the PHOENIX atmosphere model \citep[BT-SETTL;][]{2013MSAIS..24..128A},
we first computed the absolute fluxes by integrating the grid PHOENIX spectra for $T_\mathrm{eff}=3600, 
3700, 3800, 3900, 4000, 4100, 4200, 4300$ K over the $Kp-$band. 
We then performed a Monte Carlo simulation, in which $T_\mathrm{eff}$ and $R_s$ 
were randomly perturbed for both of \tarcb\ and \tarcc\ assuming Gaussian distributions
based on the values in Table \ref{hyo1}, and absolute fluxes were interpolated and converted 
into the photon count ratio between the two stars. Consequently, we found the relative flux contribution 
from \tarcb\ is $0.367\pm0.075$ while that of \tarcc\ is $0.633\pm0.075$ in the $Kp-$band.

The actual flux contribution from each star depends on which aperture we use. 
We used aperture A for the light curve fitting (Figure \ref{fig:aperture}). In order to estimate the relative
contributions from \tarcb\ and \tarcc\ for this aperture, we summed the total flux counts
in the postage stamp ($N_\mathrm{tot}$), the counts in the pixels in the upper half of 
the postage stamp which are ``not" in the aperture ($N_1$), and the counts in the pixels in the lower half of 
the postage stamp which are not in the aperture ($N_2$). 
The resulting ratios $N_1/N_\mathrm{tot}$ and $N_2/N_\mathrm{tot}$ can approximately be considered
as the relative flux ratios from \tarcb\ and \tarcc\ that are not inside the photometric aperture. 
Therefore, by subtracting these ratios from the intrinsic flux ratios above (0.367 and 0.633)
and renormalizing them, we finally obtained the relative flux contributions for aperture A
as $0.357\pm0.077$ and $0.643\pm 0.077$ for \tarcb\ and \tarcc, respectively. 
In fitting the transit light curve, we took this dilution factor into account for \tarcc.

\begin{figure*}
\centering
\includegraphics[width=18.5cm]{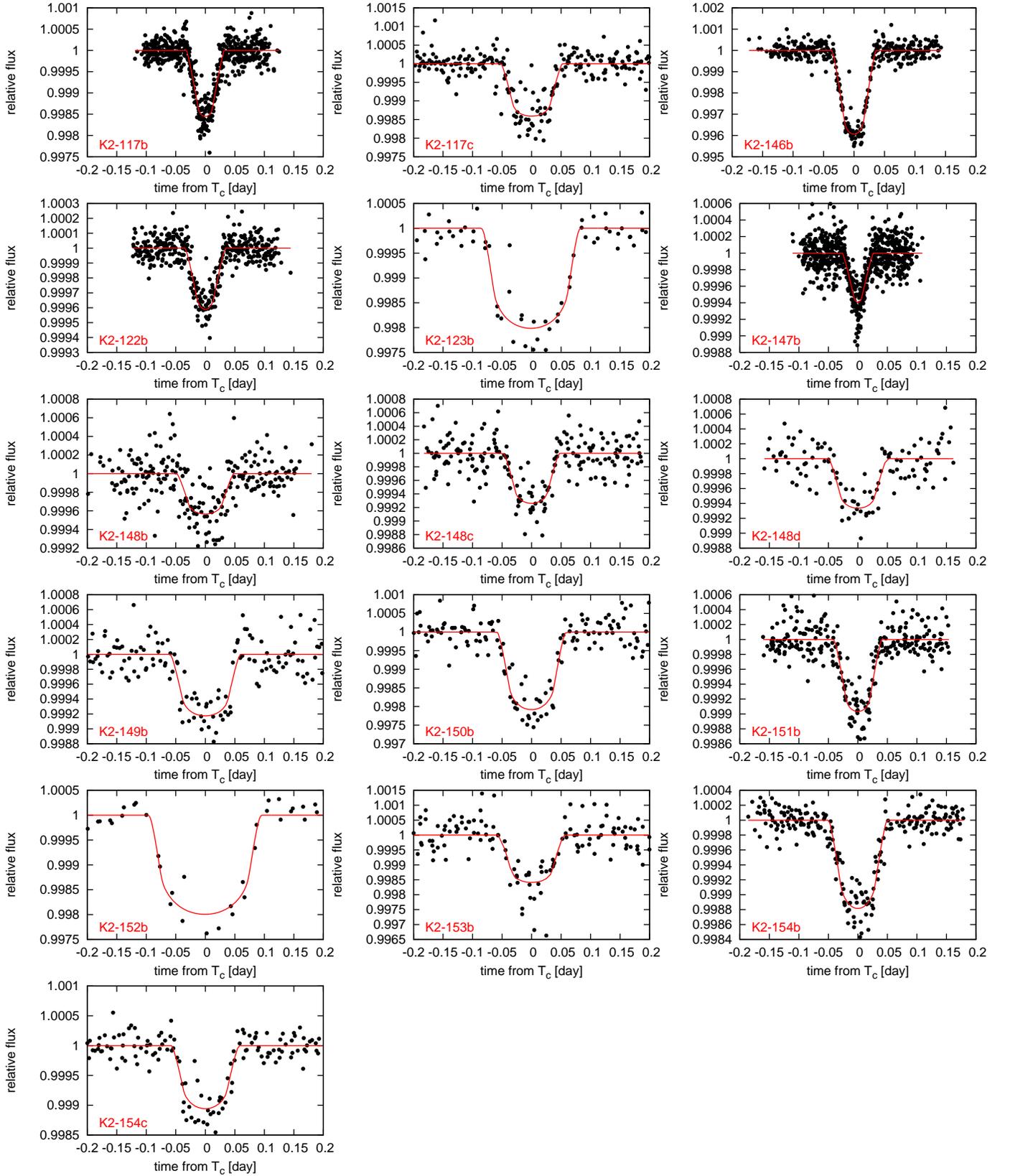}
\caption{{\it K2} light curves around transits for individual candidates folded by their periods. 
Possible TTVs are corrected and all the transits are aligned in these light curves. 
For \tarcc, the flux contamination from \tarcb\ is taken into account and the dilution factor
is corrected. The best-fit transit curves are shown by the red solid lines. }
\label{fig:folded}
\end{figure*}

\begin{table*}[tb]
\tabletypesize{\tiny}
\rotatebox{90}{
\begin{minipage}{\textheight}
\caption{Planetary Parameters}\label{hyo2}
\begin{tabular}{lcccccccc}
\hline\hline
Planet  & FPP & $P$ (days) & $T_{c,0}$ ($\mathrm{BJD}-2454833$) & $a/R_s$ & $R_p/R_s$ & $R_p$ ($R_\oplus$) & $a$ (AU) & $S$ ($S_\oplus$) \\\hline
K2-117b & $4.5\times 10^{-6}$ & $1.291563\pm 0.000026$ & $2305.90021\pm 0.00082$ & $9.4_{-0.5}^{+0.4}$ & $0.0362_{-0.0007}^{+0.0008}$ & $2.03_{-0.21}^{+0.21}$ & $0.0188\pm 0.0007$ & $123.6\pm 28.2$  \\
K2-117c & $< 10^{-6}$ & $5.44425\pm 0.00032$ & $2305.12220\pm 0.00208$ & $19.7_{-4.0}^{+2.0}$ & $0.0347_{-0.0013}^{+0.0018}$ & $1.94_{-0.21}^{+0.22}$ & $0.0491\pm 0.0017$ & $18.1\pm 4.1$  \\
K2-146b & $< 10^{-6}$ & $2.644646\pm 0.000043$ & $2306.35327\pm 0.00085$ & $15.5_{-2.5}^{+0.9}$ & $0.0577_{-0.0012}^{+0.0021}$ & $2.20_{-0.23}^{+0.23}$ & $0.0266\pm 0.0010$ & $20.7\pm 4.8$  \\
K2-122b & $1.9\times 10^{-5}$ & $2.219315\pm 0.000074$ & $2306.60981\pm 0.00125$ & $13.6_{-3.2}^{+1.3}$ & $0.0183_{-0.0007}^{+0.0017}$ & $1.22_{-0.13}^{+0.17}$ & $0.0288\pm 0.0009$ & $95.7\pm 21.5$  \\
K2-123b & $1.2\times 10^{-4}$ & $30.9542\pm 0.0022$ & $2283.53953\pm 0.00476$ & $61.6_{-15.3}^{+6.4}$ & $0.0413_{-0.0015}^{+0.0031}$ & $2.66_{-0.28}^{+0.33}$ & $0.1641\pm 0.0053$ & $2.7\pm 0.6$  \\
K2-147b & $1.0\times 10^{-4}$ & $0.961917\pm 0.000026$ & $2468.94616\pm 0.00125$ & $8.8_{-2.1}^{+1.4}$ & $0.0229_{-0.0011}^{+0.0016}$ & $1.38_{-0.15}^{+0.17}$ & $0.0159\pm 0.0005$ & $200.1\pm 45.7$  \\
K2-148b & $3.7\times 10^{-6}$ & $4.38395\pm 0.00080$ & $2557.05956\pm 0.00961$ & $16.7_{-4.5}^{+3.2}$ & $0.0193_{-0.0019}^{+0.0021}$ & $1.33_{-0.18}^{+0.19}$ & $0.0454\pm 0.0014$ & $48.8\pm 11.0$  \\
K2-148c & $5.3\times 10^{-5}$ & $6.92260\pm 0.00070$ & $2554.72777\pm 0.00458$ & $27.3_{-6.9}^{+3.6}$ & $0.0251_{-0.0018}^{+0.0025}$ & $1.73_{-0.21}^{+0.24}$ & $0.0616\pm 0.0019$ & $26.5\pm 6.0$  \\
K2-148d & $1.5\times 10^{-4}$ & $9.7579\pm 0.0010$ & $2553.34305\pm 0.00545$ & $36.3_{-9.6}^{+6.0}$ & $0.0238_{-0.0020}^{+0.0026}$ & $1.64_{-0.21}^{+0.24}$ & $0.0774\pm 0.0024$ & $16.8\pm 3.8$  \\
K2-149b & $< 10^{-6}$ & $11.3320\pm 0.0013$ & $2555.33834\pm 0.00600$ & $34.3_{-7.9}^{+3.7}$ & $0.0264_{-0.0012}^{+0.0018}$ & $1.64_{-0.18}^{+0.20}$ & $0.0830\pm 0.0027$ & $7.0\pm 1.6$  \\
K2-150b & $1.5\times 10^{-5}$ & $10.59357\pm 0.00084$ & $2558.96158\pm 0.00392$ & $32.2_{-9.5}^{+3.6}$ & $0.0420_{-0.0016}^{+0.0038}$ & $2.00_{-0.21}^{+0.27}$ & $0.0727\pm 0.0027$ & $4.9\pm 1.1$  \\
K2-151b & $1.8\times 10^{-6}$ & $3.835592\pm 0.000023$ & $2558.40166\pm 0.00104$ & $18.4_{-5.0}^{+2.1}$ & $0.0289_{-0.0010}^{+0.0019}$ & $1.35_{-0.14}^{+0.16}$ & $0.0365\pm 0.0014$ & $20.8\pm 4.8$  \\
K2-152b & $2.0\times 10^{-6}$ & $32.6527\pm 0.0035$ & $2742.96234\pm 0.00479$ & $56.9_{-13.2}^{+5.0}$ & $0.0408_{-0.0015}^{+0.0029}$ & $2.81_{-0.30}^{+0.34}$ & $0.1735\pm 0.0054$ & $2.9\pm 0.7$  \\
K2-153b & $7.3\times 10^{-5}$ & $7.51554\pm 0.00098$ & $2747.91718\pm 0.00524$ & $24.2_{-7.2}^{+3.5}$ & $0.0371_{-0.0019}^{+0.0030}$ & $2.00_{-0.22}^{+0.26}$ & $0.0601\pm 0.0021$ & $11.8\pm 2.7$  \\
K2-154b & $4.3\times 10^{-6}$ & $3.67635\pm 0.00017$ & $2748.37866\pm 0.00202$ & $13.4_{-4.7}^{+1.5}$ & $0.0315_{-0.0012}^{+0.0042}$ & $2.23_{-0.24}^{+0.37}$ & $0.0408\pm 0.0012$ & $57.5\pm 12.9$  \\
K2-154c & $1.5\times 10^{-6}$ & $7.95478\pm 0.00063$ & $2743.38098\pm 0.00350$ & $25.3_{-5.3}^{+2.2}$ & $0.0297_{-0.0012}^{+0.0019}$ & $2.10_{-0.23}^{+0.25}$ & $0.0683\pm 0.0021$ & $20.5\pm 4.6$  \\
\hline
\end{tabular}
\end{minipage}
}
\end{table*}

\begin{figure}
\centering
\includegraphics[width=8.5cm]{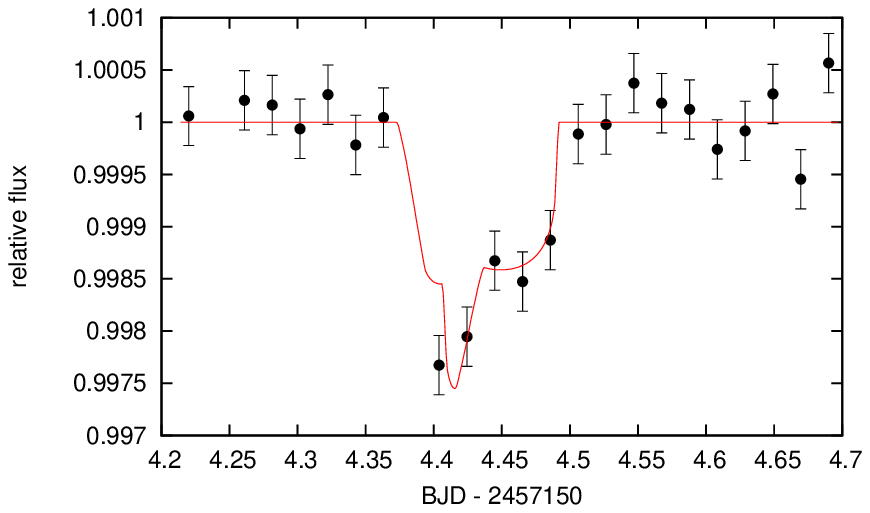}
\caption{First double transit event observed for \taraa. The best-fit model is shown by the red solid line. }
\label{fig:doublet1}
\end{figure}

\begin{figure}
\centering
\includegraphics[width=8.5cm]{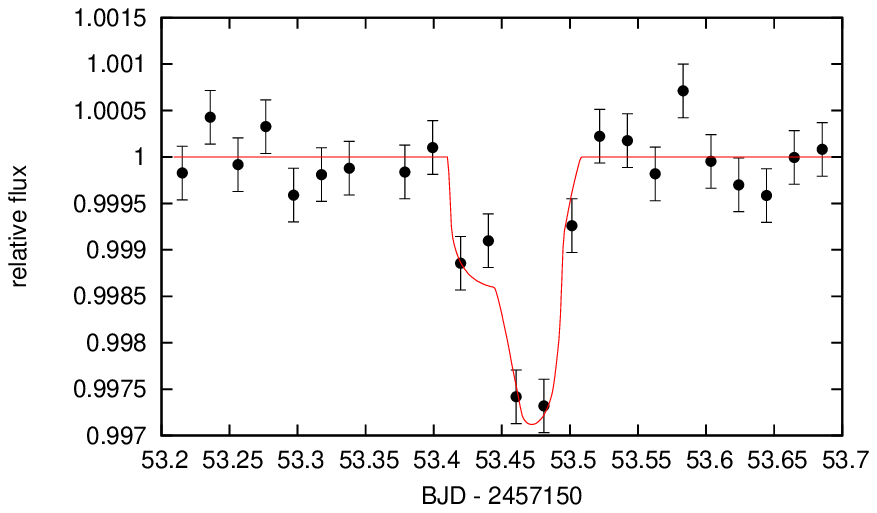}
\caption{Seond double transit event observed for \taraa. The best-fit model is shown by the red solid line. }
\label{fig:doublet2}
\end{figure}

\begin{figure}
\centering
\includegraphics[width=8.5cm]{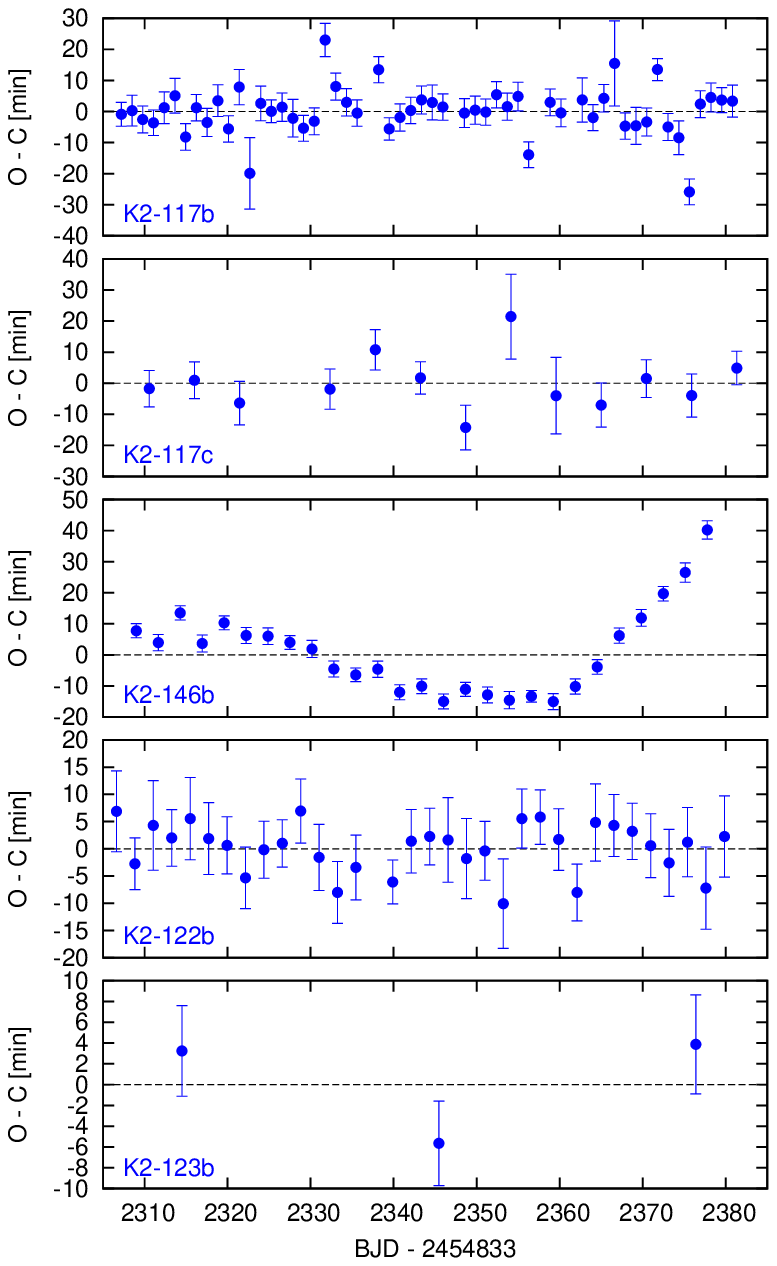}
\caption{$O-C$ diagrams for mid-transit times for {\it K2} campaign field 5 planets. }
\label{fig:f5_ttv}
\end{figure}

\begin{figure}
\centering
\includegraphics[width=8.5cm]{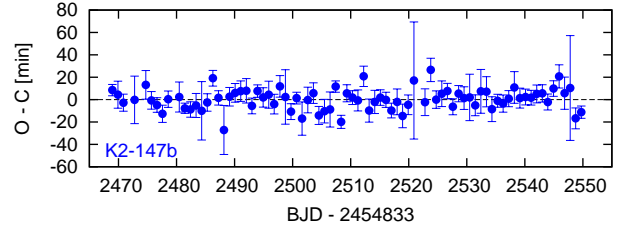}
\caption{$O-C$ diagram for mid-transit times for \tarba b.}
\label{fig:f7_ttv}
\end{figure}

\begin{figure}
\centering
\includegraphics[width=8.5cm]{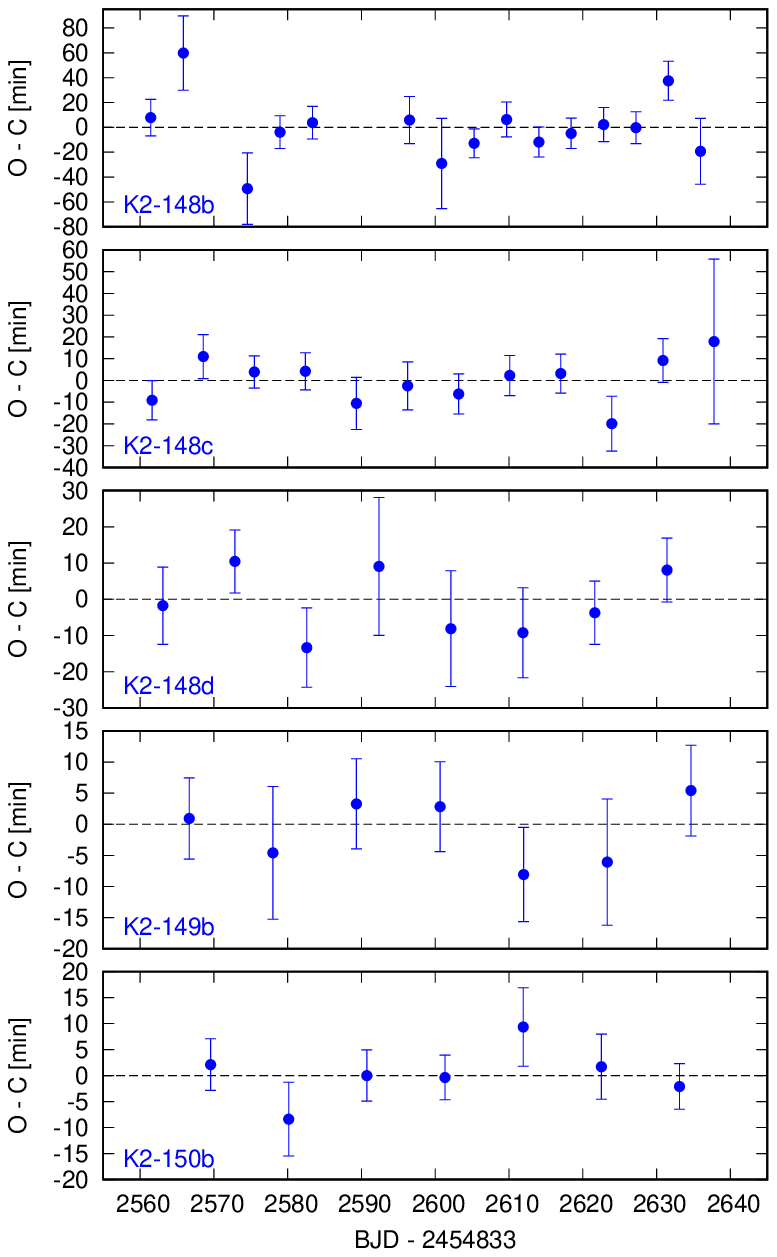}
\caption{$O-C$ diagrams for mid-transit times for {\it K2} campaign field 8 planets.}
\label{fig:f8_ttv}
\end{figure}

\begin{figure}
\centering
\includegraphics[width=8.5cm]{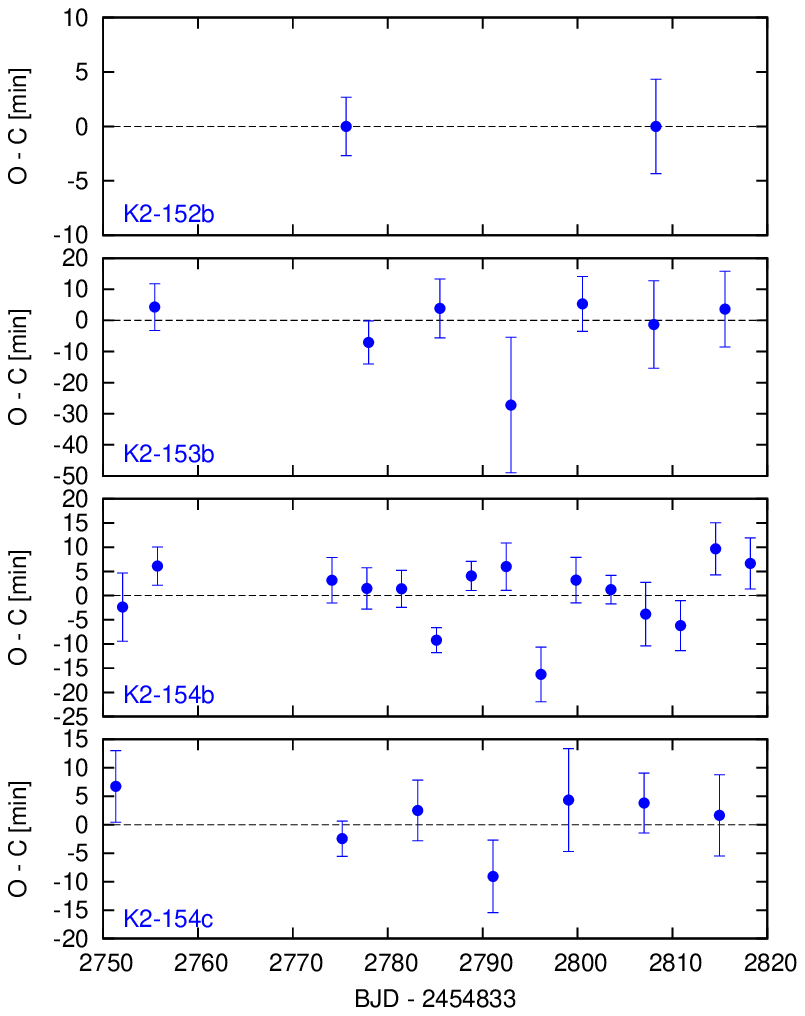}
\caption{$O-C$ diagrams for mid-transit times for {\it K2} campaign field 10 planets.}
\label{fig:f10_ttv}
\end{figure}

After fitting the light curve segments for each planet candidate, we
obtained the transit parameters summarized in Table \ref{hyo2}. Figure
\ref{fig:folded} plots the folded {\it K2} data around the transits
(black points) along with the best-fit light curve models (red solid
lines) for individual planet candidates.  For \taraa, double transit
events, in which two planets transit the host star simultaneously,
were predicted and identified in two light curve segments, and we
fitted these segments separately with only $T_c$ and baseline
coefficients floating freely (Figures \ref{fig:doublet1} and
\ref{fig:doublet2}).  Using the optimized $T_c$ datasets, we fitted
the observed $T_c$'s for each candidate with a linear ephemeris and
estimated the orbital period $P$ and transit-center zero point
$T_{c,0}$, which are also listed in Table \ref{hyo2}.  We note
  that in Figure \ref{fig:folded}, the data for some of the planet
  candidates exhibits a larger scatter in the residuals during the
  transits, compared to the data outside of transits.  This increased
  scatter during transits could be ascribed to spot-crossings for
  relatively active stars \citep[e.g.,][]{2011ApJ...743...61S}, but
  the large outliers are probably the instrumental artifacts and were
  clipped in the light curve analysis.  In order to check the
absence/presence of TTVs, we plot the observed minus calculated
($O-C$) diagrams of $T_c$ for each candidate in Figures
\ref{fig:f5_ttv}--\ref{fig:f10_ttv}.  Visual inspection suggests that
\tarab\ exhibits a strong TTV while the other candidates show no clear
sign of TTVs.  Based on the stellar and transit parameters, we also
estimate the planet radius $R_p$, semi-major axis $a$, and insolation
flux from the host star $S$, as also shown in Table \ref{hyo2}.

\subsubsection{Fitting Ground-based Transits\label{s:ground}}\label{s:ground}
Because the transit signals of \tarcf b are difficult to detect in the ground-based light curves, 
not all the transit parameters can be constrained from these light curves alone.
We therefore fitted these light curves by fixing $a/R_s$ and $b$ at the values determined 
from the {\it K2} light curves. We also fixed the limb-darkening parameters at the theoretical 
values of ($u_1, u_2$) = ($0.37, 0.40$), ($0.33, 0.41$), ($0.45, 0.12$), ($0.02, 0.37$), and 
($-0.01, 0.26$) for the $g'$, $r'$, $z_\mathrm{s}$, $J$, and $K_\mathrm{s}$ bands, respectively. 
For each transit, we fitted the multi-band data simultaneously by allowing $R_p/R_s$ for each 
band and a common $T_c$ to be free. In addition, we simultaneously modeled the baseline 
systematics adopting a parameterization introduced by \cite{2016AJ....152..171F}, which takes 
account of the second-order extinction effect. The applied function is 
\begin{eqnarray}
m_\mathrm{t} (t) = M_\mathrm{tr} + k_0 + k_t t + k_\mathrm{c} m_\mathrm{c} (t) + \Sigma k_i X_i, 
\end{eqnarray}
where $m_\mathrm{t}$ and $m_\mathrm{c}$ are the apparent magnitudes of the target star  
and comparison stars, respectively, $M_\mathrm{tr}$ is a transit model in magnitude scale, 
$t$ is time, $X_i$ is auxiliary observables such as stellar displacements on the detectors, 
sky backgrounds, and FWHM of the stellar PSFs, and $k_0$, $k_t$, $k_\mathrm{c}$, and $k_i$ a
re coefficients to be fitted. For the auxiliary observables, we included only the ones that show 
apparent correlations with the light curves;
the stellar displacements in X direction and sky backgrounds (in magnitude scale) were included 
for the $J$-band light curve and none was included for the other light curves.

\begin{figure}
\centering
\includegraphics[width=8.5cm]{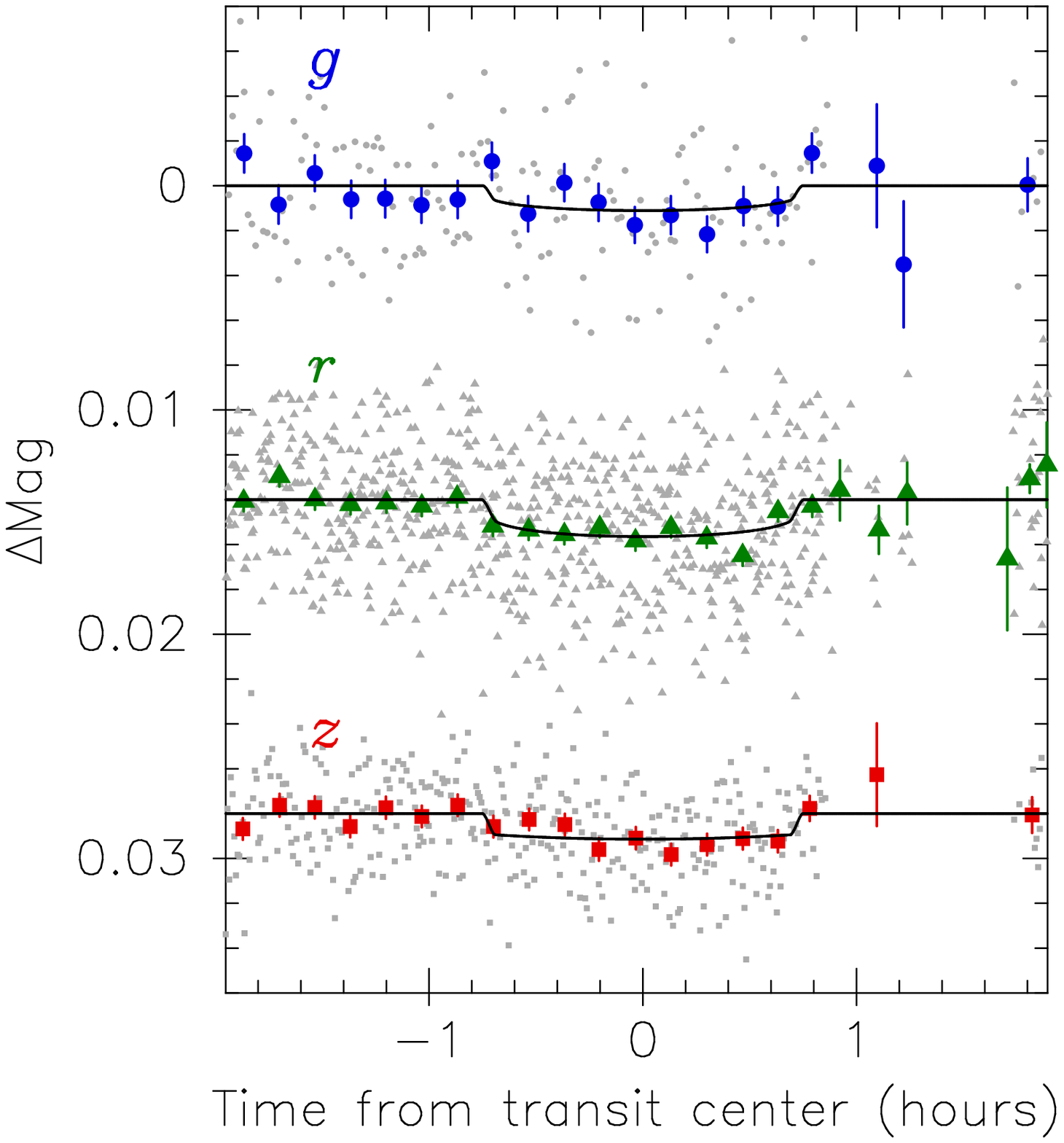}
\caption{Ground-based transit observation for \tarcf\ by OAO/MuSCAT (grey dots). The binned flux data
for $g'-$, $r'-$, and $z_\mathrm{s}-$bands are shown by the blue circles, green triangle, 
and red squares, respectively. The black solid lines indicate the best-fit transit models for 
individual bands. }
\label{fig:msct}
\end{figure}

\begin{figure}
\centering
\includegraphics[width=8.5cm]{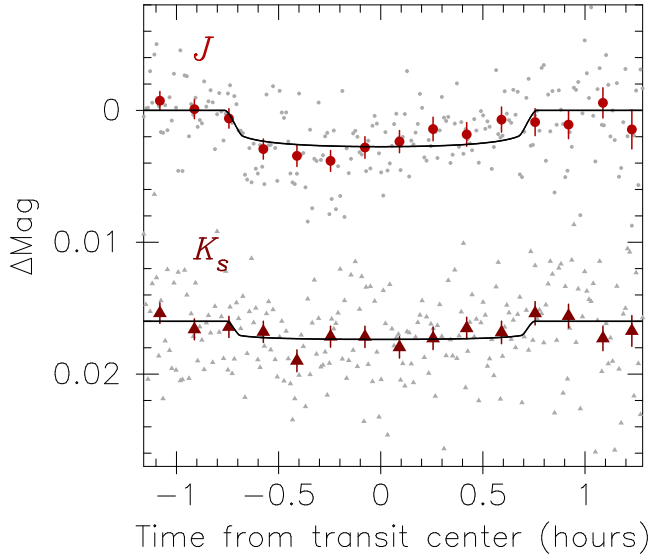}
\caption{Ground-based transit observation for \tarcf\ by IRSF/SIRIUS (grey dots). The binned flux data
for $J-$, and $K_\mathrm{s}-$bands are shown by the dark-red circles, brown triangles, 
respectively. The black solid lines indicate the best-fit transit models for 
individual bands. }
\label{fig:irsf}
\end{figure}

\begin{table}[tb]
\begin{center}
\caption{Results of Follow-up Transit Observations for \tarcf}\label{hyo3}
\begin{tabular}{lcc}
\hline
bandpass & $R_p/R_s$ & $T_c$ ($\mathrm{BJD}-2454833$) \\\hline\hline
\textit{(MuSCAT observation)} & & $2819.2215\pm0.0015$ \\ 
$g'$ & $0.0295_{-0.0098}^{+0.0070}$ & \\
$r'$ & $0.0360_{-0.0032}^{+0.0029}$ &  \\
$z_\mathrm{s}$ & $0.0312_{-0.0048}^{+0.0042}$ & \\
\textit{(SIRIUS observation)} & & $2834.5651_{-0.0017}^{+0.0013}$ \\ 
$J$ & $0.0295_{-0.0098}^{+0.0070}$ & \\
$K_s$ & $0.0360_{-0.0032}^{+0.0029}$ & \\
\hline
\end{tabular}
\end{center}
\end{table}

To obtain the best estimates and uncertainties of the free parameters, we performed 
an MCMC analysis using a custom code \citep{2013PASJ...65...27N}. We first optimized 
the free parameters using the AMOEBA algorithm \citep{1992nrca.book.....P}, and 
rescaled the error bar of each data point so that the reduced $\chi^2$ becomes unity. 
To take into account approximate time-correlated noises, we further inflated each error 
bar by a factor $\beta$, which is the ratio of the standard deviation of a binned residual 
light curve to the one expected from the unbinned residual light curve assuming white 
noises alone \citep[][]{2006MNRAS.373..231P,2008ApJ...683.1076W}. We then implemented 
10 and 50 independent MCMC runs with 10$^6$ steps each for the MuSCAT and SIRIUS 
data, respectively, and calculated the median and 16 (84) percentile values from the 
merged posterior distributions of the individual parameters. The resultant values are 
listed in Table \ref{hyo3} and the systematics-corrected light curves along with the best-fit 
transit models are shown in Figures \ref{fig:msct} and \ref{fig:irsf}.

We note that the detections of these transit signals are marginal. The $\chi^2$ improvement 
by the best-fit transit model over a null-transit one ($R_p/R_s$ are forced to be zero)  for the 
MuSCAT data is 58.7, to which 6.4, 37.8, and 14.5 are contributed from the $g'$-, $r'$-, and 
$z_\mathrm{s}$-band data, respectively, corresponding to the 6.5$\sigma$ significance 
given the number of additional free parameters of four. In the same way, the $\chi^2$ 
improvement for the SIRIUS data is 24.2, to which 15.6 and 6.6 are contributed from the 
$J$- and $K_\mathrm{s}$-band data, respectively, corresponding to the 4.2$\sigma$ 
significance given the number of additional free parameters of three. Nevertheless, as discussed 
below, all the $R_p/R_s$ values are largely consistent with each other and all the $T_c$ values 
are well aligned, both supporting that these transit detections are positive.

\begin{figure}
\centering
\includegraphics[width=8.5cm]{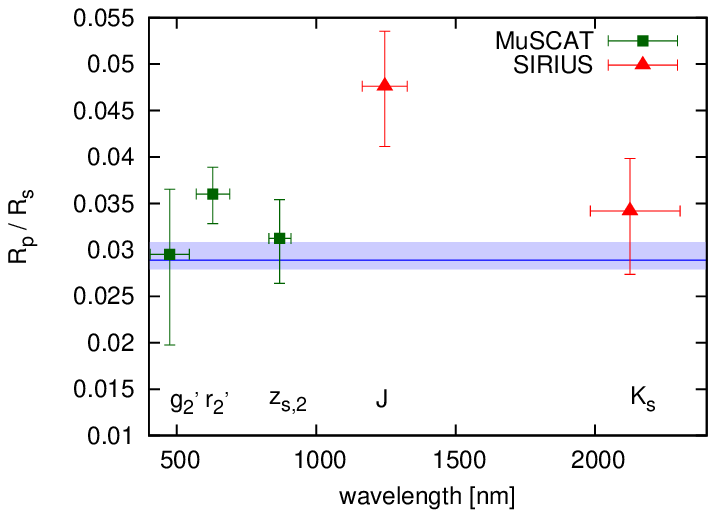}
\caption{Observed $R_p/R_s$ values of  for \tarcf b in different bandpasses. The blue horizontal line
and its upper and lower shaded areas indicate $R_p/R_s$ and its $\pm1\sigma$ errors in the $K_p$ band. }
\label{fig:epic21087_rprs}
\end{figure}
Based on the results of the ground-based transit observations, we compare the 
transit depths in different bandpasses. In Figure \ref{fig:epic21087_rprs}, 
the $R_p/R_s$ value for each band is plotted as a function of wavelength. 
The blue horizontal line indicates $R_p/R_s$ in the $Kp$ band, for which
the $\pm 1\sigma$ errors are shown by the blue shaded area. The
transit depths in the $g'$, $r'$, $z_\mathrm{s}$, and $K_s$ bands are consistent 
with the {\it K2} result within $2\sigma$, while the $J-$band result exhibits a moderate
disagreement. But as is seen in Figure \ref{fig:msct}, the $J-$band light curve
seems to suffer from a systematic flux variation, which has not been corrected
by our light-curve modeling. A more sophisticated light-curve analysis using
e.g., Gaussian processes \citep[see e.g.,][]{2015MNRAS.451..680E} may be able to settle this issue.

\begin{figure}
\centering
\includegraphics[width=8.5cm]{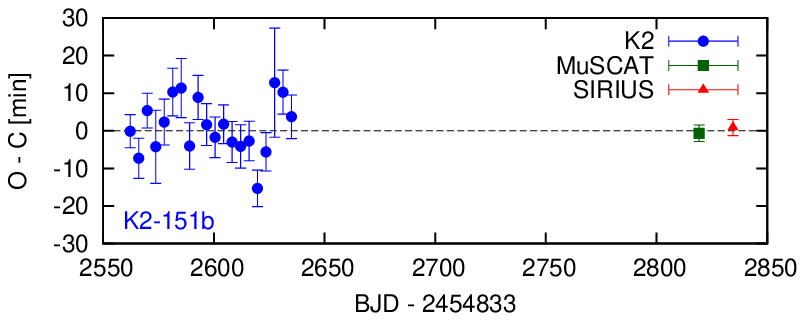}
\caption{$O-C$ diagram for mid-transit times for \tarcf b. Ground-based transit observations are
shown by the green square (MuSCAT) and red triangle (SIRIUS). }
\label{fig:epic21087_ttv}
\end{figure}
In the absence of the follow-up transit observations, we obtained the orbital
period as $P=3.83547\pm0.00015$ days from the {\it K2} data alone. 
Our ground-based transit observations were conducted $>180$ days after the {\it K2} 
observation for campaign 8 was over, as shown in Figure \ref{fig:epic21087_ttv}. 
These follow-up observations improved the precision in the orbital period of \tarcf b by a factor of $> 6$. 
Figure \ref{fig:epic21087_ttv} also implies that the mid-transit times observed by
{\it K2} are consistent with the follow-up transit observations, and no clear sign of TTV
is seen for \tarcf b.

\subsection{Validating Planets\label{s:validation}}\label{validation}

We used the open source {\tt vespa} software package \citep{2015ascl.soft03011M} to compute 
the false positive probabilities (FPPs) of each planet candidate. Similar to previous statistical validation 
frameworks \citep{2011ApJ...727...24T, 2014MNRAS.441..983D}, {\tt vespa} relies upon Galaxy model 
stellar population simulations to compute the likelihoods of both planetary and non-planetary scenarios 
given the observations. In particular, {\tt vespa} uses the {\tt TRILEGAL} Galaxy model \citep{2005A&A...436..895G} 
and considers false positive scenarios involving EBs, background EBs (BEBs), as well as 
hierarchical triple systems (HEBs). {\tt vespa} models the physical properties of the host 
star taking into account broadband photometry and spectroscopic stellar parameters using {\tt isochrones} 
\citep{2015ascl.soft03010M}, and compares a large number of simulated scenarios to the observed 
phase-folded light curve. Both the size of the photometric aperture and contrast curve constraints are 
accounted for in the calculations, as well as any other observed constraints such as the maximum depth 
of secondary eclipses allowed by the data. Finally, {\tt vespa} computes the FPP for a given planet candidate 
as the posterior probability of all non-planetary scenarios.

Inputting all available information (e.g., folded {\it K2} light curves, contrast curves from AO imaging, 
constraint on the depths of secondary eclipses, and spectroscopic parameters of the target stars) from
our follow-up observations and analyses, we ran {\tt vespa} and calculated FPP for each planet candidate. 
Table \ref{hyo2} summarizes thus derived FPP for our planet candidates; all the FPP values are
well below the fiducial criterion of planet validation ($\mathrm{FPP}<1\%$), by which 
the planet candidates listed in Table \ref{hyo2} are statistically validated.

AO observations by Subaru/IRCS and HiCIAO allowed us to obtain high resolution images 
of candidate planet hosts, but our imaging can only cover the FoV of $\sim 20^{\prime\prime}\times 20^{\prime\prime}$. 
Moreover, the targets were not imaged at the exact center of the detector, 
and nearby stars within {\it K2} photometric apertures may be missing in our high resolution images. 
In order to ensure that such missing stars are not sources of 
false positive (i.e., BEBs), we checked the archived catalogs \citep[e.g.,][]{2005yCat.1297....0Z, 2012ApJS..203...21A}
to look for faint nearby sources for each target. 
As a consequence, we found that \tarab, \tarba, \tarcc, and \tarce\ have nearby faint stars,
which could be inside the {\it K2} photometric apertures ($\sim 30^{\prime\prime}\times 30^{\prime\prime}$)
\footnote{Here, the faint star around \tarab\ is different from the two faint 
sources that we identified in the HiCIAO image. The faint nearby source around \tarcc\ is also 
different from \tarcb.}. 
The delta magnitudes of these nearby stars are larger than $\Delta m_r=5$ mag, 
but smaller than those corresponding to the observed transit depths. Among the four systems, however, 
the nearby stars around \tarab, \tarcc, and \tarce\ are located around the edge 
of the {\it K2} photometric apertures (separation larger than $10^{\prime\prime}$), and so a significant 
fraction of light from those faint stars should be missing in the {\it K2} photometry ($>40\%$). 
Given this loss of light, 
we found it almost impossible to account for the observed transit depths even for the maximum 
occultation case (i.e., $50\%$ loss of light during eclipses).

Concerning \tarba, we identified two faint sources around the target, which are separated
by $10\farcs5$ ($\Delta m_R=6.1$ mag) and $10\farcs8$ ($\Delta m_R=6.7$ mag), 
respectively. Given the observed transit depth of $\sim 0.06\%$, either of these faint stars could 
be the source of the observed signal. To prove that this is not the case, we created new {\it K2} 
light curves using customized apertures for this object, which excluded the pixels around those 
faint stars. This analysis revealed that the transits are indeed
reproduced even after excluding these faint stars, by which we concluded that 
\tarba\ is the source of transits.

Finally, we checked if the stellar densities estimated via transit fitting are consistent with
the spectroscopically estimated densities, in order to make sure that the planets are 
indeed transiting the low-mass host stars. As a result, we found that the stellar densities from 
the transit modeling all have super-solar densities, suggesting that the 
planets are transiting low-mass stars, and are in good agreement with spectroscopic values 
within $1\sigma$ except \taraa b, for which the two densities are compatible within $2\sigma$. 
Based on all these facts above as well as the {\tt vespa} calculations and absence of large 
RV variations for a fraction of systems, we conclude that the candidates in Table \ref{hyo2} are 
all bona-fide planets \footnote{We note that false positives of an instrumental origin are
very unlikely, since our candidates do not include one whose period is close to the known
periods associated with instrumental artifacts (e.g., the 6-hour rolling motion).}.

\section{Individual Systems \label{s:individual}}\label{s:individual}

\subsection{\taraa}

The planet candidate \taraa b ($P=1.29$ days, $R_p=2.03R_\oplus$) was first reported 
by \citet{2016MNRAS.461.3399P} and recently \citet{2017AJ....154..207D} validated this candidate 
along with the additional planet \taraa c of similar size ($R_p=1.94R_\oplus$), orbiting 
the same star with $P=5.44$ days. We report here independent validations of these planets
using our own observational data (AO and a high resolution spectrum), and have performed a more 
thorough analysis, including the double transit modeling (Figures \ref{fig:doublet1} and \ref{fig:doublet2}) 
and TTV analysis. As shown in Figure \ref{fig:f5_ttv}, 
no clear TTV signals are seen in the $O-C$ diagram. 
The two planets exhibit moderate transit depths ($\sim 0.15\%$), enabling transit follow-up observations 
from the ground, by which we can refine transit parameters and ephemerides. 

\subsection{\tarab}

\tarab\ is the coolest star in our sample, for which we obtain $T_\mathrm{eff}=3385$ K. 
\citet{2016MNRAS.461.3399P} and \citet{2017AJ....154..207D} reported that \tarab\
hosts a mini-Neptune candidate in a $2.645-$day orbit with a possible TTV. 
We have performed a global fit to the {\it K2} light curve allowing every transit center to float 
freely, and confirmed the TTV as shown in Figure \ref{fig:f5_ttv}. As a result of 
inputting the TTV-corrected transit curve to \texttt{vespa}, we were able to validate
\tarab b as a bona-fide planet. The strong TTV ($>30$ minutes) suggests that
the object causing TTV is either a very massive planet or has an orbit very close to the mean
motion resonance (MMR), although the detailed TTV modeling is beyond the scope 
of this paper. 

\tarab\ also exhibits the deepest transit among our sampled stars, making it a very
unique target for atmospheric characterizations and TTV modeling by transit follow-ups
from the ground and space. However, the predicted transit times are now highly uncertain 
due to the TTV combined with the long time interval
after the {\it K2} observation, and it would be required to cover a long baseline around 
predicted transits. Fortunately, \tarab\ is supposed to be observed by {\it K2} again 
in the Campaign field 16, by which we can refine the ephemeris and possibly 
put a constraint on the object inducing the TTV. 

\tarab\ is very faint in the optical ($m_V=16.2$ mag), but given 
the magnitudes in the near infrared (e.g., $m_H=11.6$ mag) one may be able to 
constrain the masses of \tarab b and the additional body by RV measurements with upcoming 
near infrared spectrographs \citep[e.g., IRD;][]{2014SPIE.9147E..14K}. 
Adopting the empirical mass-radius relation for small planets by \citet{2014ApJ...783L...6W}, 
the mass of \tarab b is estimated as $\sim 5.6M_\oplus$ and the corresponding
RV semi-amplitude induced by this planet is $\sim 5.1$ m s$^{-1}$. 

\subsection{\tarad}

\tarad\ is a quite metal-rich early M dwarf ($\mathrm{[Fe/H]}=0.37\pm0.12$), 
hosting a close-in Earth-like planet ($R_p=1.22R_\oplus$, $P=2.22$ days). 
\citet{2016MNRAS.461.3399P} reported this system to be a candidate planet-host, which
was later validated by \citet{2017AJ....154..207D}. In addition to an independent validation
by AO imaging and high resolution spectroscopy, we attempted a measurement of the planet mass. 
As shown in Figure \ref{fig:rv}, however, RVs measured by FIES and HARPS-N show
a small variation. Assuming a circular orbit, we fit the observed RV datasets, for which we find the 
RV semi-amplitude of $K=-2.6\pm 4.5$ m s$^{-1}$.
This is consistent with a non-detection, but the $1\sigma$ upper limit of $K$
translates to $\approx 2.9M_\oplus$ for \tarad b's mass, suggesting that its composition may be
somewhat similar to that of the Earth. Future monitoring with a greater number of RV points
would allow for a more robust mass measurement.

\subsection{\tarac}

The detection of a transiting mini-Neptune ($R_p=2.66R_\oplus$) was reported around \tarac\ by 
\citet{2016MNRAS.461.3399P}, and \citet{2017AJ....154..207D} later validated this planet. 
We have presented our own observations and data analysis including the precise RV measurement
(Figure \ref{fig:rv}), and independently validated \tarac b as a genuine planet in a $31-$day orbit. 

The relatively large orbital distance ($a=0.164$ AU) translates to \tarac b's equilibrium temperature
of $325$ K on the assumption that its Bond albedo is 0.3 ($\sim $ Earth's albedo). Thus, the planet is near the
potential habitable zone, making it an attractive target for further characterizations. 
Given the moderate transit depth ($\sim 0.2\%$), the detection of transits is relatively
easy with 2-m class ground telescopes, but one may have a small chance to 
observe a complete transit due to the long orbital period. 

\subsection{\tarba}

\tarba\ is a metal-rich M dwarf, 
orbited by a super-Earth with the ultra-short period (USP; $\sim 23$ hours). 
No detection has so far been reported for this planet. 
According to exoplanet.eu\footnote{http://exoplanet.eu/catalog}, \tarba b is the seventh validated USP planet ($P<1$ day)
around M dwarfs after Kepler-32f, Kepler-42c, Kepler-732c, KOI-1843.03 \citep{2013ApJ...773L..15R}, K2-22b 
\citep{2015ApJ...812..112S}, and K2-137b \citep{2018MNRAS.474.5523S}. 
Interestingly, these planets show an increasing trend in $R_p$ as a function of the orbital 
period $P$. We will later discuss the dependence of planetary sizes on insolation flux
from host stars. 

\subsection{\tarca}

The transit-like signal was first detected for this target with a period of $17.09$ days
and we measured its depth and duration as $0.245\%$ and $1.64$ hours. 
As shown in Figures \ref{fig:contrast} and \ref{fig:ccf}, however, \tarca\
is comprised of at least two stars separated by $\sim 0\farcs3$. 
The transit curve is also V-shaped, and the preliminary light-curve fitting preferred a grazing transit. 
We thus conclude that either of the two stars seen in Figure \ref{fig:contrast}
has an eclipsing stellar companion (a late M dwarf or a brown dwarf), which is responsible
for the relative Doppler shift in the cross-correlation profile (Figure \ref{fig:ccf}). 
Indeed, as we described in Section \ref{s:hires}, multiple spectra were obtained for this target
by Subaru/HDS with the I$_2$ cell but the RV analysis did not converge, which is most likely because
the observed spectra (with the I$_2$ cell) for RV measurement are different in shape from 
the template (without the I$_2$ cell), which complicates the fitting procedure.

In Figure \ref{fig:ccf}, the two line positions in the cross-correlation profile are separated by 
$\Delta \mathrm{RV}= 18$ km s$^{-1}$. 
The template spectrum for \tarca\ was taken 
at $\mathrm{JD}=2457676.037$, which corresponds to the orbital phase of $\phi\sim 0.19$
when folded by the period of \tarca.01. This phase implies that the left line ($\mathrm{RV}\sim 19$ km s$^{-1}$) 
in the cross-correlation profile corresponds to the star with a companion (i.e., EB) and 
right one ($\mathrm{RV}\sim 37$ km s$^{-1}$) corresponds to the other star. Assuming a circular orbit 
($e=0$) and the orbital inclination 
of $90^\circ$ for the EB, we can roughly estimate the secondary-to-primary mass ratio $q$ via
\begin{eqnarray}
\label{eq:K}
\Delta \mathrm{RV}=212.9083\left(\frac{M_\mathrm{1}/M_\odot}{P/\mathrm{day}}\right)^{\frac{1}{3}}
\frac{q}{(1+q)^{\frac{2}{3}}}\sin\phi~~(\mathrm{km~s}^{-1}),~~~
\end{eqnarray}
where $M_1$ is the mass of the primary star. When we adopt $M_1=0.6M_\odot$, we obtain
$\sim 0.2M_\odot$ for the mass of the secondary. This would be easily confirmed by 
taking additional spectra for the absolute RV measurement. 
\tarca\ provides a good testing bench, where high resolution imaging and/or high dispersion 
spectroscopy become powerful tools to identify and characterize hierarchical triple systems.

\subsection{\tarcb\ and \tarcc}

As we have seen in Section \ref{s:fitK2}, \tarcc\ turned out to host three planets,
whose radii we estimate as $1.33R_\oplus$, $1.73R_\oplus$, and $1.64R_\oplus$ for the innermost ($P=4.38$ days),
middle ($P=6.92$ days), and outermost ($P=9.76$ days) planets, respectively. 
In order to see if \tarcb\ and \tarcc\ are bound to each other (common proper-motion stars), 
we checked the proper motions of the two stars and found $(\mu_\alpha, \mu_\delta)=
(-34.9\pm6.8~\mathrm{mas~yr^{-1}}, -27.3\pm7.7~\mathrm{mas~yr^{-1}})$
and $(-38.4\pm9.4~\mathrm{mas~yr^{-1}}, -26.7\pm3.1~\mathrm{mas~yr^{-1}})$,
for \tarcb\ and \tarcc, respectively \citep{2013MNRAS.433.2054S}, 
indicating that the two stars share the same proper motion within the errorbars. 
The almost identical RV values (Figure \ref{fig:ccf}), along with the same distance (Table \ref{hyo1}) to the stars,
all imply that \tarcb\ and \tarcc\ are bound to each other. 
The separation of $9\farcs4$ between the stars translates to the projected distance 
of $\sim 1100$ AU from each other. 
It is of interest that one of the two late-type stars in a wide binary orbit has 
multiple super-Earths. Searching for planets around \tarcb\ also helps us understand the planet 
formation in cool wide-binary systems. 

The period ratio of \tarcc b and c is close to the 2:3 MMR. 
We investigated possible TTVs for the three planets, but no clear signal is seen in 
Figure \ref{fig:f8_ttv}, likely due to the small planetary masses.

\subsection{\tarcd}

\tarcd\ is a slightly metal-rich early M dwarf, having a super-Earth ($R_p=1.6R_\oplus$) 
in a 11-day orbit. The RV measurement by Subaru/HDS shows no significant RV variation, 
supporting the planetary nature of \tarcd b.

\subsection{\tarce}

The validated super-Earth \tarce b is similar to \tarcd b in terms of its 
period ($P=11$ days) and size ($R_p=2.0R_\oplus$), except that 
it is orbiting a cooler host star ($T_\mathrm{eff}=3499$ K). 
Two absolute RVs were measured by Subaru/HDS, which are consistent
within their errors. Given the moderate-depth transit ($\sim 0.2\%$) for a super-Earth,
\tarce\ is a good target for ground-based transit observations to refine
system parameters and search for a possible TTV.

\subsection{\tarcf}

\tarcf\ is a metal-poor M dwarf hosting a transiting small planet with $P=3.84$ days. 
The size of \tarcf b ($R_p=1.35R_\oplus$) suggests that 
it is likely a rocky planet. 
The relative brightness of the host star allowed us to observe the follow-up transits
from the ground, enabling a considerable improvement in the transit ephemeris (Section \ref{s:ground}). 
We also measured rough RVs, which completely ruled out the EB scenario. 
\tarcf\ is also a good target for future precise RV measurements in the near infrared;
with $m_J=10.93$ mag, new and upcoming spectrographs like IRD and
CARMENES \citep{2014SPIE.9147E..1FQ} may be able to pin down the mass of \tarcf b. 

\subsection{\tarda}

The transiting mini-Neptune \tarda\ is orbiting the host M dwarf every 33 days. 
Assuming the Bond albedo of $A_B=0.3$, we estimate the equilibrium temperature of
\tarda b as $T_\mathrm{eq}=331$ K, putting this planet near the habitable zone. 
The host star's brightness ($m_V=13.73$ mag and $m_J=10.96$ mag) and moderate 
transit depth ($\sim 0.2\%$) make this system a good target for further follow-ups
including precise RV measurements, either in visible and near infrared, and ground-based 
transit observations. 
Based on the mass-radius relation by \citet{2014ApJ...783L...6W}, the mass of
\tarda b is $\sim 7.0M_\oplus$, corresponding to the RV semi-amplitude of
$K\sim 1.9$ m s$^{-1}$. 

\subsection{\tardb}

We did not obtain multiple spectra for \tardb, 
which does not allow us to rule out completely the grazing EB scenario. 
Our HDS spectrum for \tardb, however,
was taken at $\mathrm{JD}=2457920.857$ corresponding to $\phi\sim 0.23$, around
which we expect to see the largest line separation in the spectrum if the transit signal
is caused by an EB. 
We carefully inspected the secondary line in the cross-correlation profile, but found no 
evidence, supporting the result of the \texttt{vespa} validation. 
\tardb\ is a slightly metal-poor, early-to-mid M dwarf orbited by a super-Earth 
($R_p=2.0R_\oplus$) with $P=7.5$ days. 

\subsection{\tardc}

We identified and validated two transiting mini-Neptunes ($R_p=2.23R_\oplus$ and $2.10R_\oplus$) around \tardc, 
a slightly metal-rich early M dwarf. The orbital periods are 3.68 and 7.95 days for
\tardc b and c, respectively, whose ratio is somewhat close to the 2:1 resonance. 
We searched for TTVs for this system, but found no clear evidence as shown 
in Figure \ref{fig:f10_ttv}. A longer-term transit follow-ups with a better
$T_c$ precision would be required.

\section{Discussion\label{s:discussion}}\label{s:discussion}

All together, we have validated 16 planets around 12 of the low-mass stars observed
by {\it K2}, based on high-resolution imaging and optical spectroscopy.
Since the number of planets around M dwarfs has been increasing rapidly, thanks to {\it K2} and other
projects, it is tempting to investigate the entire ensemble of M-dwarf planets, seeking patterns
among their properties.  We 
focus here on a search for any relationships between planet size, the stellar insolation (the flux received by the planet),
and the stellar metallicity.  This is because insolation and metallicity are strongly suspected of
playing an important role in the formation and evolution of planets, and some possible
correlations with planetary radius have already been discussed in the literature 
\citep[e.g.,][]{2013ApJ...775..105O, 2014Natur.509..593B, 2015MNRAS.453.1471D, 2016NatCo...711201L}.

\begin{table}[tb]
\begin{center}
\caption{Revised Spectroscopic Parameters Based on \texttt{SpecMatch-Emp}}\label{hyo4}
\begin{tabular}{lccc}
\hline
System & $T_\mathrm{eff}$ (K) & [Fe/H] (dex) & $R_s$ ($R_\odot$) \\\hline\hline
K2-3 & $3799\pm70$ & $-0.25\pm0.12$ & $0.500\pm0.050$ \\
K2-5 & $4056\pm70$ & $-0.44\pm0.12$ & $0.607\pm0.061$ \\
K2-9 & $3502\pm70$ & $-0.43\pm0.12$ & $0.358\pm0.036$ \\
K2-18 & $3463\pm70$ & $0.01\pm0.12$ & $0.427\pm0.043$ \\
K2-26 & $3680\pm70$ & $-0.06\pm0.12$ & $0.504\pm0.050$ \\
K2-54 & $4012\pm70$ & $-0.18\pm0.12$ & $0.630\pm0.063$ \\
K2-72 & $3393\pm70$ & $-0.49\pm0.12$ & $0.370\pm0.037$ \\
K2-83 & $3806\pm70$ & $-0.05\pm0.12$ & $0.565\pm0.057$ \\
\hline
\end{tabular}
\end{center}
\end{table}

To this end, we created a list of transiting planets around M dwarfs based on 
information in the NASA Exoplanet Archive\footnote{https://exoplanetarchive.ipac.caltech.edu}, 
exoplanet.eu, and exoplanets.org\footnote{http://exoplanets.org}.
We restricted our sample to confirmed or validated planets around dwarf stars
with $T_\mathrm{eff}\leq 4000$ K. We excluded unvalidated planet candidates. 
We also excluded 6 systems for which spectroscopic characterization is not available
(Kepler-1350, 1582, 1624, 1628, 1646, and 1649). 

For some systems, different investigators have reported different values for stellar and planetary parameters,
sometimes differing by more than $3\sigma$. For the sake of homogeneity,
we adopted the stellar parameters of \citet{2013ApJ...779..188M, 2013AJ....145...52M,
2016ApJ...818...46M, 2016AJ....152...61M, 2017AJ....153...64M, 2017AJ....153..267M} 
for a majority of the {\it Kepler} and {\it K2} stars in our sample, since those were derived based on the same (or similar)
observing and reduction schemes.
We also used the \texttt{SpecMatch-Emp} code to derive our own versions of the stellar
parameters (Table \ref{hyo4}), for cases in which high-resolution spectra were available on the
ExoFOP website\footnote{https://exofop.ipac.caltech.edu}.
As noted by \citet{2017ApJ...836...77Y}, the M-dwarf parameters 
derived by the \texttt{SpecMatch-Emp} code were calibrated using the sample of \citet{2015ApJ...804...64M}, 
faciliating comparisons.
For the other systems, for which high-resolution spectra were not available, 
we adopted the stellar parameters from the literature 
\citep{2012ApJ...748...93R, 2014MNRAS.443.1810B, 2015ApJ...800...99T, 2015AJ....149..166H, 2015Natur.527..204B, 
2016ApJ...820...41H, 2017ApJ...836..167D, 2017ApJ...837...72M, 2017Natur.542..456G, 2017Natur.544..333D}, 
although no metallicity values were reported by \citet{2017ApJ...837...72M}. 
Planet radii were estimated based on the revised stellar radii and
the values of $R_p/R_s$ reported in the literature or by the \textit{Kepler} team. 

\begin{figure}
\includegraphics[width=8.5cm]{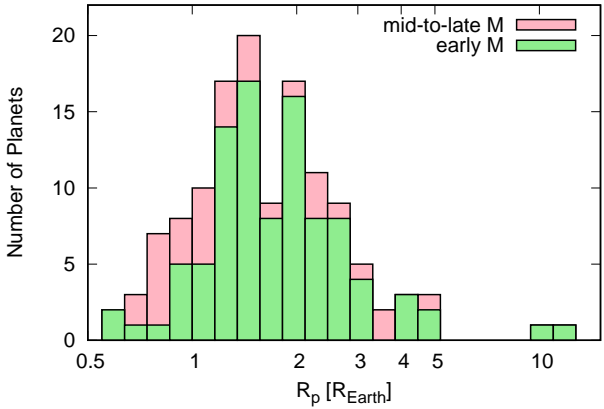}
\caption{Histogram of planet radius, for the validated and well-characterized transiting planets
around M dwarfs. The number counts for mid-to-late M dwarfs are shown above those for early M dwarfs. 
}
\label{fig:histogram}
\end{figure}
We split the sample into (1) planets around early M dwarfs (3500-4000\,K)
and (2) mid-to-late M dwarfs ($<$3500\,K), to check for any differences in planet
properties associated with stellar mass or effective temperature.
By this definition our sample consists of 96 planets around 63 early M dwarfs,
and 32 planets around 17 mid-to-late M dwarfs.

Figure \ref{fig:histogram} shows the distribution of planet sizes, on a logarithmic scale.
A larger number of Earth-sized planets ($0.5-1.25\,R_\oplus$) are found around
the later-type stars, in spite of the smaller number of such stars in our sample.
Although no completeness correction has been applied, it is interesting that 
Figure~\ref{fig:histogram} shows that both types of stars have deficit of planets 
with $R_p=1.57-1.82\,R_\oplus$, relative to somewhat smaller or larger planets.
This is consistent with the findings of \citet{2017AJ....154..109F} 
and \citet{2017arXiv171005398V}, 
based mainly on solar-type stars, that planets with sizes between 1.5-2~$R_\oplus$ are rarer 
than somewhat smaller or larger planets.
This paucity has been interpreted as the outcome of photoevaporation on a population of 
planets with rocky cores ($\approx1.5\,R_\oplus$) with differing masses of gaseous envelopes 
and different levels of irradiation \citep{2017ApJ...847...29O}, 
or as the outcome of the erosion of planetary envelopes by internal heat from 
cooling rocky cores \citep{2017arXiv170801621G}. 
The same sort of deficit seen in Figure~\ref{fig:histogram} suggests
that the same processes seem to be taking place around M dwarfs.

\subsection{Insolation Dependence}

\begin{figure}
\centering
\includegraphics[width=8.5cm]{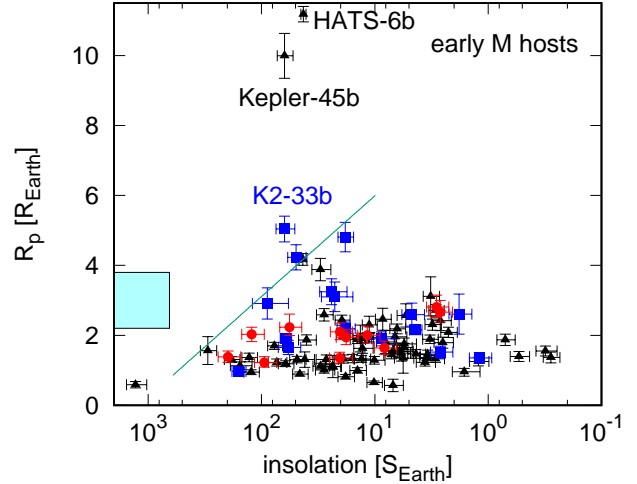}
\caption{Stellar insolation fluxes vs. radii of planets around early M dwarfs 
($3500~\mathrm{K}<T_\mathrm{eff}\leq4000~\mathrm{K}$).
Our newly validated planets (red circles), other planets discovered by {\it K2} (blue squares), and planets from 
the {\it Kepler} primary mission and other surveys (black triangles).
The cyan rectangle area is the ``hot-super-Earth desert" described by
\citet{2016NatCo...711201L}. 
See the text for the upper boundary of $R_p$ (green solid line). 
}
\label{fig:insolation_earlyM}
\end{figure}

\begin{figure}
\centering
\includegraphics[width=8.5cm]{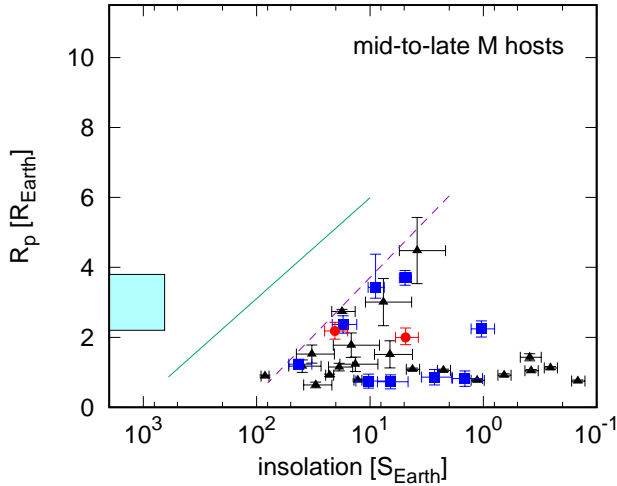}
\caption{Stellar insolation fluxes vs. radii of planets around mid-to-late M dwarfs 
($T_\mathrm{eff}\leq 3500~\mathrm{K}$).
Symbols and plot ranges are the same as in Figure \ref{fig:insolation_earlyM}. 
}
\label{fig:insolation_midM}
\end{figure}

Figures \ref{fig:insolation_earlyM} and \ref{fig:insolation_midM} display the planet radius
as a function of stellar insolation $S$. 
In these figures, red circles represent our newly validated planets,
blue squares are other {\it K2} planets, and black triangles are
planets discovered during the primary \textit{Kepler} prime mission or by ground-based surveys.
Looking at Figures \ref{fig:insolation_earlyM} and \ref{fig:insolation_midM}, we note that an important 
contribution of {\it K2} has been the discovery of relatively large planets ($R_p\gtrsim 2.5R_\oplus$), 
which were not frequently detected during the \textit{Kepler} primary mission.


Figures \ref{fig:insolation_earlyM} and \ref{fig:insolation_midM} show
a lack of larger planets ($R_p\gtrsim 2\,R_\oplus$) in the close proximity of M stars. 
The deficit of close-in planets ($P\lesssim 2$ days) was previously reported by, e.g.,
\citet{2012ApJS..201...15H, 2016A&A...589A..75M, 2017AJ....154..109F} mainly for 
solar-type stars.
In order to draw a rough boundary above which planets are apparently rare,
we took an approach similar to that described in \citet{2016MNRAS.461.1841C} for the
planet-mass vs. stellar-metallicity diagram. Namely, we computed the
cumulative weighted distribution of $R_p$ for each insolation bin with its width 
being $0.2$ in the $\log S$ space\footnote{The bin size was set to 0.1 in 
$\log S$, and thus each bin is overlapping with the neighboring bins}. 
We then estimated the maximum radius for each bin by finding the $97\,\%$ upper limit 
of this cumulative distribution. Finally, these upper limits were fitted with a linear
function in the $\log S-R_p$ space. We restricted this analysis to close-in planets
($P\lesssim 10$ days) and excluded hot Jupiters ($R_p>8R_\oplus$)
since they seem to form a different population from their smaller counterparts
\citep[e.g.,][]{2016A&A...589A..75M}.

The green line in Figure \ref{fig:insolation_earlyM} represents this estimated
boundary line. The moderate slope of the line 
($R_p/R_\oplus=(-2.88\pm 0.47)\log S/S_\oplus +(8.87\pm 0.91)$) implies that 
only larger planets ($R_p\gtrsim 3R_\oplus$) are missing in the proximity of the host stars. 
\citet{2013ApJ...775..105O} showed that close-in low-mass planets are likely to suffer significant
envelope evaporation due to the X-ray and extreme ultraviolet (EUV) radiation from the host star. 
On the other hand, theoretical works have shown that the gravitational potential 
of hot Jupiters is so deep that the XUV radiation from host stars cannot significantly
strip their envelopes \citep[e.g.,][]{2009ApJ...693...23M}, which is consistent 
with the presence of the few hot Jupiters seen in Figure \ref{fig:insolation_earlyM}. 
\citet{2013ApJ...775..105O} also noted that the evaporation of hydrogen envelopes
should occur within the first 100~Myr, when stars are at the peak of their chromospheric activity.
In this light, it is interesting that K2-33b seems to be unusually large for its
level of current irradiation; the host is a pre-main-sequence star with an age of $\approx\,11$ Myr.
This suggests that K2-33b is actively evaporating, and that
its radius will shrink significantly over the next 100~Myr.
Note that we did not exclude K2-33b from the analysis to draw the boundary.

The cyan rectangles in Figures~\ref{fig:insolation_earlyM}
and~\ref{fig:insolation_midM} depict the ``hot-super-Earth" desert
discussed by \citet{2016NatCo...711201L}, for close-in planets around
solar-type stars (i.e., $2.2\,R_\oplus<R_p<3.8\,R_\oplus$ and
$S>650\,S_\oplus$).  Evidently this rectangle is not a good description
of the ``desert'' seen around M dwarfs.  Instead, for M dwarfs, the
``desert'' seems to extend towards much lower insolation. 
Also interesting is that the observed ``desert'' is shifted toward
lower insolation for the mid-to-late M stars. 
In Figure~\ref{fig:insolation_midM}, we draw a similar upper boundary of $R_p$ 
for the mid-to-late M sample by the purple dashed line. 
The derived slope of this boundary ($R_p/R_\oplus=(-3.34\pm 0.34)\log S/S_\oplus +(7.05\pm 0.42)$)
agrees with that for the early-M sample to within $1\,\sigma$. 
To make this easier to see, the same green line that was drawn in Figure~\ref{fig:insolation_earlyM} 
is also drawn in Figure~\ref{fig:insolation_midM}.

This result can be understood in the framework of
\citet{2013ApJ...775..105O}, which implies that plotting the planet
radius against the current bolometric insolation is not the most
direct way to seek evidence for photoevaporation.  Envelope
evaporation is caused specifically by X-ray and EUV irradiation from
the star, and not by the bolometric flux.  This is especially so for M
dwarfs because they emit a higher fraction of X-rays relative to the
bolometric flux than solar-type stars.  Thus planets around M dwarfs
should have been eroded more efficiently, relative to planets around
solar-type stars with the same level of bolometric insolation. This
was shown in Figure 7 of \citet{2013ApJ...775..105O}, wherein the lack
of large planets extends to smaller bolometric fluxes for later-type
stars.  \citet{2013ApJ...775..105O} also showed that when $R_p$ is
plotted against the empirically estimated X-ray exposure, the maximum
planet size at a given X-ray exposure is approximately the same for
all types of host stars. Although we do not attempt here to reproduce
this type of plot, a comparison between Figures
\ref{fig:insolation_earlyM} and \ref{fig:insolation_midM} does suggest
a similar pattern.
We note that this pattern is also compatible with the scenario in which photoevaporation 
is responsible for the radius gap (Figure \ref{fig:histogram}), and favors photoevaporation
over planetary internal heat as the explanation \citep{2017arXiv170801621G},
because in the latter case it should be the bolometric luminosity (not the XUV luminosity) that is
relevant to atmospheric loss.

Another possible mechanism that could lead to a deficiency of close-in 
planets with large sizes is high-eccentricity migration \citep[e.g.,][]{1996Sci...274..954R, 2011ApJ...742...72N} 
coupled with the disruption of planetary envelopes in the vicinity of the Roche limit 
\citep{2016ApJ...820L...8M, 2017AJ....154..192G}. 
Since Neptune-sized planets are often observed to have lower mean densities than
Jovian or Earth-sized planets, their planetary envelopes should be relatively easy to strip.
\citet{2015ApJ...814..130M} and \citet{2017ApJ...842...40L} suggested that the decline 
of planet occurrence rate of all sizes at shortest orbital distances ($P<10$ days) 
could be the result of disk truncation at these orbital distances. 
Several mechanisms that truncate the planet populations around different types of
stars are discussed in the literature \citep[e.g.,][]{2013ApJ...769...86P, 2015ApJ...814..130M},
including tidal halting of migrating planets. 
The lack of planets of all sizes at higher insolation level in Figures \ref{fig:insolation_earlyM} 
and \ref{fig:insolation_midM} may also be consistent this interpretation. In this picture, the disk 
truncation likely happens at $\approx2$-day period for both early and mid-to-late M dwarfs to 
explain the lack of detected planets. 
However, the ``truncation" we observed is not a vertical boundary in the insolation vs. radius plane 
as one would expect in the disk truncation picture, instead it has a moderate slope. 
In other words, at high insolation levels, there is only a lack of larger planets but 
not smaller planets.  This would seem to favor 
the photoevaporation picture rather than the disk truncation picture.

Figures \ref{fig:insolation_earlyM} and \ref{fig:insolation_midM} also suggest
a lack of large planets at low insolations (i.e., at longer orbital periods; $P\gtrsim 10$ days). 
This could be related to the formation process of these larger planets, which
somehow is easier in their observed locations; the two figures illustrate that large planets 
including the hot Jupiters ($R_p\gtrsim 3\,R_\oplus$) seem to occur within a relatively
narrow range of periods. 
However, given that the occurrence rate of planets with $R_p>3\,R_\oplus$ is known to 
dwindle dramatically and long-period planets are more affected by detection biases 
associated with the transit geometry and short span of the {\it K2} monitoring, 
it is premature to draw any conclusions on those outer planets. 
Compared to planetary systems around solar-type stars,
little is known on the formation and evolution of M-dwarf planets, but
measurements of eccentricity for close-in planets and other orbital parameters 
(e.g., the stellar obliquity) would help to test all these hypotheses 
for M-dwarf planets.

\subsection{Metallicity Dependence}

Stellar metallicity is also known to be related to planet size
in exoplanetary systems \citep[see, e.g.,][]{2014Natur.509..593B}. 
It is well known that the occurrence rate of giant planets around solar-type
stars depends sensitively on [Fe/H] \citep[e.g.,][]{2010PASP..122..905J}. 
The occurrence of Earth and Neptune-sized planets were reported to be less dependent 
on metallicity \citep[e.g.,][]{2008A&A...487..373S, 2011arXiv1109.2497M}, although some recent studies 
have shown that such planets are at least somewhat more frequent around metal-rich 
solar-type stars \citep[e.g.,][]{2015AJ....149...14W}. 
In particular, there is growing evidence that small close-in planets ($P<10$ days) are 
preferentially found around metal-rich stars 
\citep{2016AJ....152..187M, 2017arXiv170607807D, 2017arXiv171204042P}. 
Specifically, \citet{2017arXiv171201198W} derived the critical period, below which
small planets orbit statistically metal-rich host stars ($P_\mathrm{crit}\approx 8.3$ days).

\begin{figure}
\centering
\includegraphics[width=8.5cm]{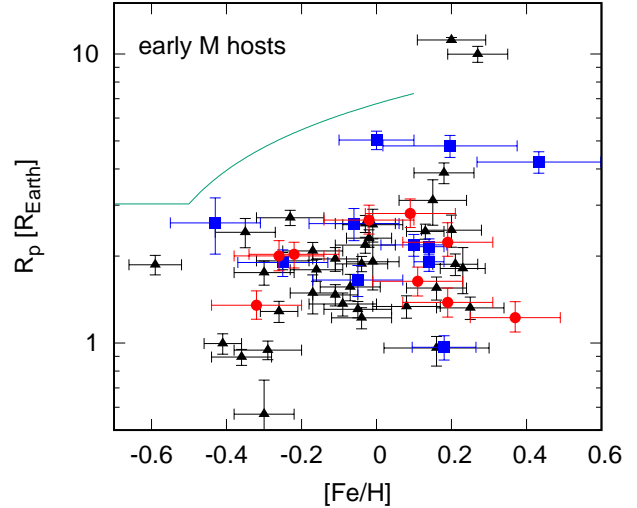}
\caption{Host stars' metallicities from spectroscopy vs. radii of the planets around 
early M dwarfs ($3500~\mathrm{K}<T_\mathrm{eff}\leq4000~\mathrm{K}$).
For multi-planet systems, the largest planets are plotted. 
Symbols are the same as in Figure \ref{fig:insolation_earlyM}. 
Note that contrary to Figures \ref{fig:insolation_earlyM} and \ref{fig:insolation_midM}, 
the $y-$scale is logarithmic. 
}
\label{fig:metallicity_earlyM}
\end{figure}

\begin{figure}
\centering
\includegraphics[width=8.5cm]{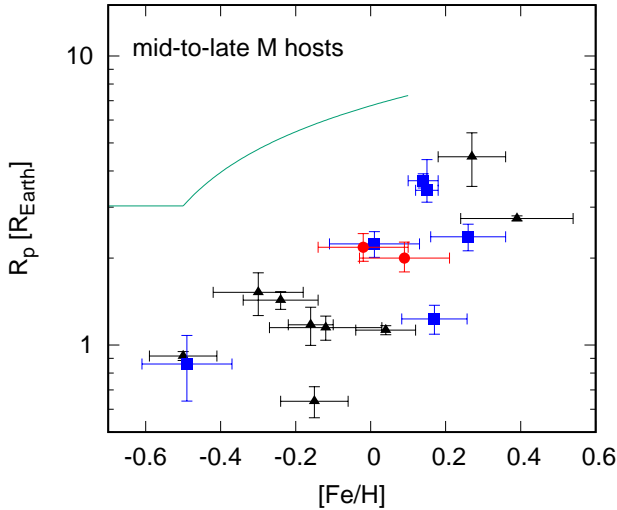}
\caption{Host stars' metallicities from spectroscopy vs. radii of the planets around mid-to-late M dwarfs 
($T_\mathrm{eff}\leq 3500~\mathrm{K}$).
For multi-planet systems, the largest planets are plotted. 
Symbols are the same as in Figure \ref{fig:insolation_earlyM}.
}
\label{fig:metallicity_midM}
\end{figure}

Here we examine the relationship between $R_p$ and [Fe/H] for M-dwarf planets,
based on our new measurements and the parameters available in the literature. 
Previously, \citet{2010A&A...519A.105S} found a hint that
planet-hosting M dwarfs are preferentially found in the region of the $(m_V-m_{K_s})-M_{K_s}$
diagram where one expects metal-rich stars to be located.
\citet{2012ApJ...748...93R} also investigated the metallicity of eight 
planet-hosting M dwarfs. They found that M-dwarf planets appear to be hosted
by systematically metal-rich stars, and that Jovian planet hosts are more
metal rich than Neptune-sized planet hosts.
\citet{2012ApJ...753...90M}, however,
found no significant difference in $g-r$ color, a metallicity indicator, between the 
planet-candidate cool hosts and other cool stars. They ascribed the apparently 
high metallicity of cool planet-host stars reported in the literature to contamination
of the sample by misidentified giant stars.

Figures \ref{fig:metallicity_earlyM} and \ref{fig:metallicity_midM}
show the radii of confirmed and validated transiting planets as a
function of stellar metallicity, for early-M hosts (3500-4000\,K) and
mid-to-late M hosts ($<$3500\,K).  We restricted the sample to stars
with spectroscopic measurements of [Fe/H].  For multi-planet systems,
we have plotted only the largest planet.  From these figures we see
that larger planets ($\gtrsim$3~$R_\oplus$) have only been found
around metal-rich stars ($\mathrm{[Fe/H]}\gtrsim 0.0$).  This is
similar to the situation with solar-type stars \citep{2012Natur.486..375B}. 
Moreover, the mid-to-late M dwarfs seem
to show a trend of increasing planet size with metallicity.  For early
M dwarfs the correlation (if any) is not obvious; there are many small
planets ($R_p\lesssim 2R_\oplus$) around super-solar metallicity
stars. However, it must be remembered that these results have not been
corrected for survey sensitivity.  Transit surveys have a strong bias
favoring the detection of short-period planets; there may be
larger-radius planets that have been missed due to their longer periods. 
It is most significant that there are no detections of super-Neptune planets 
around metal-poor M dwarfs (the upper left region in both figures), since such 
large planets are easier to detect than smaller planets.

Based on RV mass measurements for small planets around solar-type stars, 
it has been demonstrated that the observed maximum planet mass increases with metallicity 
\citep{2016MNRAS.461.1841C, 2017AJ....153..142P}. A similar trend is seen for
planet radius in Figures \ref{fig:metallicity_earlyM} and \ref{fig:metallicity_midM}. 
To compare the previous finding with the distribution of M-dwarf planets, we draw 
in Figures \ref{fig:metallicity_earlyM} and \ref{fig:metallicity_midM} the upper envelope by the green solid 
line corresponding to Equation (1) of \citet{2016MNRAS.461.1841C}, 
where the planet mass is converted into radius assuming $R_p/R_\oplus\propto (M_p/M_\oplus)^{0.59}$ 
\citep{2017ApJ...834...17C}; all the planets except hot Jupiters are below this line. 
Although the number of systems plotted is much smaller than in previous
works for solar-type stars, the upper envelopes of planet radius seem to be pushed
towards lower values for coolest stars.

\citet{2015MNRAS.453.1471D} advanced an explanation for the paucity of
gaseous planets around metal-poor stars. They argued that metal-rich
stars possessed protoplanetary disks with a higher surface density of solids,
which led to more rapid formation of rocky cores with a critical mass ($>2M_\oplus$) for 
gas accretion.  If the formation timescale of
critical-mass cores is longer than the disk lifetime, gaseous
planets are unlikely to form.  Although their argument focused on
planets around solar-type stars, Figures \ref{fig:metallicity_earlyM}
and \ref{fig:metallicity_midM} suggest that a similar argument might apply
to low-mass stars.

To be more quantitative, we computed the Pearson's correlation coefficient $r$
between $R_p$ and [Fe/H]. We found
$r=0.332$ and $0.689$ for early M and mid-to-late M stars, respectively,
corresponding to the $p-$values of $0.0115$ and $0.0022$.
This is evidence for some kind of relationship between planet radius
and stellar metallicity for cool stars, as
has been previously reported for solar-type stars \citep{2014Natur.509..593B}. 
The mid-to-late M dwarf sample shows a higher correlation coefficient
than that of the early M sample, but the number of the systems is also much smaller,
which may have led to an apparently higher correlation by chance. 
To check whether the two samples are drawn from the same $\mathrm{[Fe/H]}-R_p$ 
distribution, we performed a Monte Carlo simulation in which 17 systems (the number 
of mid-to-late M systems) are randomly selected from the 57 early M dwarfs, and we 
computed the probability that the correlation coefficient $r$ for the subset of 17 systems 
is higher than 0.689 (the observed $r$ for the mid-to-late M stars). 
We found that its probability is $0.0063$, implying that the mid-to-late M 
dwarf sample indeed shows a stronger correlation between $\mathrm{[Fe/H]}$
and $R_p$.

Since the envelopes of close-in planets may have been evaporated (at
least to some degree) by X-ray and EUV radiation from the star, we
also tried to compute the correlation coefficients after removing
planets for which the insolation exceeds 100 times the Earth's
insolation, approximately the minimum value for which Figure~\ref{fig:insolation_earlyM} suggests
that shrinkage takes place.
We obtained a slightly higher correlation coefficient ($r=0.352$) for the early-M sample, 
but with an almost identical statistical significance ($p=0.0114$), probably due to the smaller 
sample size.
The $R_p-$[Fe/H] correlation is especially strong for coolest M dwarfs
($T_\mathrm{eff}\leq 3500~\mathrm{K}$), suggesting that the amount of
initial solid material is extremely sensitive to the formation of Neptunian (and jovian) planets
with hydrogen-helium envelopes around coolest stars.

Another relevant factor that affects the $\mathrm{[Fe/H]}-R_p$ relation is the 
correlation between the planet period and its host star's metallicity. 
\citet{2016AJ....152..187M} and \citet{2017arXiv170607807D}
have recently shown that stars with close-in rocky planets ($P<10$ days) are preferentially
seen around metal-rich stars, and thus the $\mathrm{[Fe/H]}-R_p$ correlation could be
in part affected by the $\mathrm{[Fe/H]}-P$ correlation. 
In order to examine such a correlation for M-dwarf planets, we split the whole sample
(both early M and mid-to-late M samples) into
inner planets ($P<7$ days) and outer planets ($P>7$ days), by which the
two subsamples have approximately the same numbers of planets, and compared their
mean metallicities. Consequently, we found a slightly higher mean metallicity 
for the inner-planet subsample ($\mathrm{[Fe/H]}=-0.033\pm 0.031$) than that for the outer-planet
subsample ($\mathrm{[Fe/H]}=-0.084\pm 0.025$), but in a statistically insignificant manner
($\approx 1.3\,\sigma$ difference). More planets are needed to confirm
the $\mathrm{[Fe/H]}-P$ correlation.

Following \citet{2014Natur.509..593B}, we also computed the mean metallicity 
for our samples.
We found the weighted mean metallicity to be $\mathrm{[Fe/H]}=-0.037\pm0.010$ 
for early M dwarfs, and $0.047\pm0.017$ for mid-to-late M dwarfs. 
\citet{2010A&A...519A.105S} noted that the mean metallicity of M dwarfs 
in the solar neighborhood is $\mathrm{[Fe/H]}\approx -0.17$. 
Therefore, our result also indicates that the confirmed/validated planet-hosting M dwarfs have 
systematically high metallicities, 
The difference in the mean metallicities was also seen by \citet{2012ApJ...748...93R}, 
but here we have extended their argument down to lower-mass stars
and have used a larger number of well-characterized systems.
We note, however, that unknown selection effects and/or different methodologies for
metallicity measurements may have introduced biases in the mean metallicities in 
the two samples. Homogeneous measurements for volume-limited samples
would be required to draw a firm conclusion.

There is no obvious reason why transit surveys should have a detection
bias favoring high stellar metallicity, but there might be some
effects. For instance, since M dwarfs with higher metallicity are more
luminous than lower-metallicity counterparts for a given temperature,
it may be somewhat easier to detect planet candidates and conduct
follow-up observations for high-metallicity stars, leading to the
validation the transiting planets, as we have done in the present
paper.  Given that we have included a variety of transiting planets
detected by many space-based and ground-based surveys, it is not
straightforward to account for any detection biases associated with
stellar metallicity.  We leave this for future work.

\section{Conclusions\label{s:conclusion}}\label{s:conclusion}

As a part of our {\it K2} follow-up program \citep[e.g.,][]{2015ApJ...812..112S}, 
we have detected tens of planet candidates around M dwarfs in {\it K2} campaign fields
5--10, and conducted follow-up observations for candidate planets
around M dwarfs.  We have validated 16 transiting planets around 12
low-mass stars, out of which 12 are newly validated planets. 
All the validated planets are relatively small in size (Earth-sized to mini-Neptunes),
with periods ranging from 0.96 to 33 days.
We have also identified a hierarchical triple system (\tarca) based on AO imaging and
high resolution spectroscopy.

We also reviewed the relationships between planet size, insolation,
and metallicity that are emerging from the growing sample of M-dwarf
planets. The planet-radius distribution suggested the same ``gap'' at 
around 1.5-2~$R_\oplus$ that was found by
\citet{2017AJ....154..109F} for a larger sample of mainly solar-type stars. 
We saw an indication of
the ``desert'' of very hot planets
larger than about $2R_\oplus$, although for the coolest M stars the
desert begins at significantly lower insolation levels than for solar-type stars.
We also confirmed that planets larger than about 3~$R_\oplus$
 are preferentially seen around metal-rich
stars ($\mathrm{[Fe/H]}>0$). Moreover, we found that the
statistical significance of this trend is higher for the coolest M dwarfs.
It will be important to try and corroborate these findings with a larger sample 
and after considering selection biases.

Fortunately, \textit{the Transiting Exoplanet Survey Satellite} \citep[\textit{TESS};][]{2015JATIS...1a4003R}
will be launched and start the transit survey in the near future, 
which would make it more straightforward to deal with selection biases and 
extract the true distributions of stellar and planetary parameters with 
a larger number of sampled stars. To corroborate our findings, homogeneous 
characterizations of the systems with and without planets are essential.

Some of the new M-dwarf planets
offer excellent prospects for further characterization, including Doppler mass measurement
with optical or near-infrared spectroscopy \citep[e.g.,][]{2014SPIE.9147E..14K}. 
As discussed above, the sizes of M-dwarf planets show some qualitative trends similar
to those around solar-type stars, but they also exhibit quantitatively different
dependences on stellar insolation and metallicity. 
Perhaps the mass-radius relation for M-dwarf planets 
will also be seen to be different from that of planets around solar-type stars \citep{2014ApJ...783L...6W}. 
Measurements of orbital eccentricity and stellar obliquity could also
provide helpful clues to the processes of planet formation
and evolution around low-mass stars.

\acknowledgments 

This paper is based on data collected at Subaru Telescope, which is operated by the National 
Astronomical Observatory of Japan, on observations collected at the
Centro Astron\'omico Hispano Alem\'an (CAHA) at Calar Alto, operated jointly
by the Max--Planck Institut f\"ur Astronomie and the Instituto de Astrof\'isica
de Andaluc\'ia, and on observations obtained \emph{a}) with the Nordic 
Optical Telescope (NOT), operated on the island of La Palma jointly by Denmark, Finland, Iceland, 
Norway, and Sweden, in the Spanish Observatorio del Roque de los Muchachos (ORM) of the 
Instituto de Astrof\'isica de Canarias (IAC); \emph{b})  with the Italian Telescopio Nazionale Galileo 
(TNG) operated on the island of La Palma by the Fundaci\'on Galileo Galilei of the INAF (Istituto 
Nazionale di Astrofisica) at the Spanish Observatorio del Roque de los Muchachos of the Instituto 
de Astrofisica de Canaria.
The data analysis was in part carried out on common use data analysis computer system 
at the Astronomy Data Center, ADC, of the National Astronomical Observatory of Japan. 
We thank Akito Tajitsu, Joanna Bulger, and Ji Hoon Kim, the support astronomers at Subaru, and
Jun Hashimoto, Shoya Kamiaka, Yohei Koizumi, and Shota Sasaki
for their helps to carry out the Subaru observations. 
We also thank Santos Pedraz for carrying out the CAFOS observations at the Calar Alto observatory. 
We are very grateful to the NOT and TNG staff members for their unique and superb support during 
the observations. 
T.H.\ is grateful to Samuel Yee for providing instructions to install \texttt{SpecMatch-Emp}. 
We are thankful to Christophe Lovis, who provided the numerical mask for the spectral 
cross-correlation analysis. 
The discussions with Eric Gaidos, Hiroyuki Kurokawa, Jose Caballero, Alexis Klutsch, and
Kento Masuda were very fruitful. 
This work was supported by Japan Society for Promotion of Science (JSPS) KAKENHI Grant Number JP16K17660.
D.\,G. gratefully acknowledges the financial support of the \emph{Programma Giovani Ricercatori -- Rita Levi Montalcini -- Rientro dei Cervelli (2012)} awarded by the Italian Ministry of Education, Universities and Research (MIUR). The research leading to these results has received funding from the European Union Seventh Framework Programme (FP7/2013-2016) under grant agreement No. 312430 (OPTICON).
D.M. acknowledge financial support from the
Universidad Complutense de Madrid (UCM), the Spanish Ministry of Economy
and Competitiveness (MINECO) from project AYA2016-79425-C3-1-P.
I.R.\ acknowledges support by the Spanish Ministry of Economy and Competitiveness (MINECO) and the Fondo Europeo de Desarrollo Regional (FEDER) through grant ESP2016-80435-C2-1-R, as well as the support of the Generalitat de Catalunya/CERCA programme.
We acknowledge the very significant cultural role and reverence that the
summit of Mauna Kea has always had within the indigenous people in Hawai'i. 

\software{IRAF \citep{1986SPIE..627..733T, 1993ASPC...52..173T}, 
ACORNS pipeline \citep{2013ApJ...764..183B}, 
SpecMatch-Emp \citep{2017ApJ...836...77Y},
EVEREST \citep{2016AJ....152..100L, 2017arXiv170205488L},
PHOENIX \citep{2013MSAIS..24..128A}, 
vespa \citep{2015ascl.soft03011M},
pyfits \citep{2012ascl.soft07009B}
}




\end{document}